\documentclass{iopart} 
\usepackage{amssymb}
\usepackage{iopams} 
\usepackage{upgreek}
\usepackage{bm}
\usepackage{graphicx}
\eqnobysec
\newcommand{\sfrac}[2]{{\textstyle \frac{#1}{#2}}} 
\newcommand{\E}{{\cal E}} 
\newcommand{\B}{{\cal B}} 
\newcommand{\EE}[1]{{\cal E}^{\scriptscriptstyle \sf #1}} 
\newcommand{\dEE}[1]{\dot{\cal E}^{\scriptscriptstyle \sf #1}} 
\newcommand{\ddEE}[1]{\ddot{\cal E}^{\scriptscriptstyle \sf #1}} 
\newcommand{\BB}[1]{{\cal B}^{\scriptscriptstyle \sf #1}} 
\newcommand{\dBB}[1]{\dot{\cal B}^{\scriptscriptstyle \sf #1}} 
\newcommand{\ddBB}[1]{\ddot{\cal B}^{\scriptscriptstyle \sf #1}} 
\begin{document}
\title{Intrinsic and extrinsic geometries of a tidally deformed black hole}     
\author{Ian Vega, Eric Poisson, and Ryan Massey} 
\address{Department of Physics, University of Guelph, Guelph, Ontario,
N1G 2W1, Canada} 
\date{May 15, 2011} 
\begin{abstract} 
A description of the event horizon of a perturbed Schwarzschild black
hole is provided in terms of the intrinsic and extrinsic geometries of
the null hypersurface. This description relies on a Gauss-Codazzi
theory of null hypersurfaces embedded in spacetime, which extends the
standard theory of spacelike and timelike hypersurfaces involving the
first and second fundamental forms. We show that the 
intrinsic geometry of the event horizon is invariant under a
reparameterization of the null generators, and that the extrinsic
geometry depends on the parameterization. Stated differently, we show
that while the extrinsic geometry depends on the choice of gauge, the
intrinsic geometry is gauge invariant. We apply the formalism to
solutions to the vacuum field equations that describe a tidally
deformed black hole. In a first instance we consider a slowly-varying,
quadrupolar tidal field imposed on the black hole, and in a second
instance we examine the tide raised during a close parabolic encounter
between the black hole and a small orbiting body.   
\end{abstract} 
\pacs{04.20.-q, 04.25.Nx,04.70.-s, 04.70.Bw, 97.60.Lf}
%\maketitle

\section{Introduction} 
\label{sec:intro} 

The tidal dynamics of inspiralling compact binaries (involving neutron
stars and/or black holes) has been the subject of vigourous
investigation in the last several years, motived by the exciting
prospect of measuring tidal signatures in the gravitational waves
emitted by such systems. Some of this work has focused on calculating 
the influence of the tidal coupling on the gravitational waves, and
estimating the accuracy with which the tidal deformation of each body
can be measured \cite{flanagan-hinderer:08,  hinderer:08, read-etal:09,
  hinderer-etal:10, pannarale-etal:11}. Some has focused on
calculating the tidal deformation of neutron stars in the
post-Newtonian approximation to general relativity \cite{mora-will:04,
  berti-iyer-will:08, vines-flanagan-hinderer:11} and in the full
theory \cite{hinderer-etal:10, damour-nagar:09,
  binnington-poisson:09}. And some has focused on the tidal
deformation of nonrotating black holes \cite{poisson:04d, poisson:05,
  fang-lovelace:05, damour-lecian:09, poisson-vlasov:10}.    

An issue that is central to all these investigations is the dependence
of adopted measures of tidal deformation on the coordinates employed
to describe the spacetime geometry. In the case of neutron
stars, the coordinate independence of the {\it relativistic Love
numbers} which measure the tidal deformation of the body's external
gravitational field was firmly established by Damour and Nagar
\cite{damour-nagar:09} and Binnington and Poisson
\cite{binnington-poisson:09}. In the case of nonrotating black  
holes, however, these gauge-invariant Love numbers were shown to
vanish \cite{binnington-poisson:09}, and the identification of
nonvanishing, coordinate-independent measures of tidal deformation has
remained an open problem. For example, Poisson and Vlasov
\cite{poisson-vlasov:10} rely on light-cone coordinates to describe
the geometry of a deformed black hole, while Damour and Lecian
\cite{damour-lecian:09} rely on Weyl coordinates in a context of
stationary and axisymmetric tides. Our main objective with this paper
is to remedy this situation by providing a complete description of the
intrinsic and extrinsic geometries of a tidally deformed event
horizon, and fully clarifying the coordinate dependence of all horizon
quantities. In particular, we introduce meaningful and practical
measures of the tidal deformation of an event horizon.    

The central assumptions in our work are that the unperturbed black
hole is nonrotating and described by the Schwarzschild solution to the  
Einstein field equations, and that the tidal deformation is
sufficiently small that it can be described accurately to first order   
in a perturbative treatment. Otherwise our formulation is completely
general: the tide can be either static, slowly varying, or fully
dynamical, and there is no requirement that it be axisymmetric. Our
description of a tidally deformed event horizon relies on two major
theoretical foundations. The first is a Gauss-Codazzi theory of null
hypersurfaces embedded in spacetime, an extension of the standard
theory of (spacelike and timelike) hypersurfaces formulated in terms
of first and second fundamental forms. This material is developed here
{\it ab initio}, in spite of the fact that similar formalisms are
extant in the literature (for example, in
Refs.~\cite{dautcourt:67, jezierski-kijowski-czuchry:02, 
gourgoulhon-jaramillo:06}); our version is presented in a form
directly suited to our application to perturbed event horizons. The
second foundation is a covariant and gauge-invariant formulation of 
black-hole perturbation theory, as summarized in the work of Martel
and Poisson \cite{martel-poisson:05}.           

Our description of a null hypersurface embedded in spacetime is tied
to its generators, the congruence of null geodesics that trace the
hypersurface. We label each generator with two comoving coordinates
$\alpha^A = (\alpha,\beta)$ (with the index $A$ running over the
values 2 and 3), and we let $\lambda$ be a running parameter on each
generator. The hypersurface is charted with the intrinsic coordinates 
$(\lambda,\alpha^A)$, and its (degenerate) intrinsic geometry is fully 
characterized by the explicitly two-dimensional metric $\gamma_{AB}$, 
the analogue of the first fundamental form of a (spacelike or
timelike) hypersurface. The extrinsic geometry, on the
other hand, is characterized by a scalar $\kappa$ (a generalization of
the black hole's surface gravity), a vector $\omega_A$, and a tensor
${\cal K}_{AB}$; these are analogous to the second fundamental form of
a (spacelike or timelike) hypersurface. We examine how these
quantities transform under reparameterizations $\lambda \to
\bar{\lambda}(\lambda,\alpha^A)$ of the generators,
and show that while the extrinsic geometry of the null hypersurface
depends on the parameterization, the intrinsic geometry is
independent of the parameterization. When applied to an event horizon,
this observation becomes one of the central results of this paper: 
{\it the intrinsic geometry of a black-hole horizon is invariant under
a reparameterization of the horizon's null generators.} This statement 
implies that any measure of tidal deformation that derives from the 
induced metric $\gamma_{AB}$ is necessarily invariant under
reparameterizations. 

This result can be restated in terms of gauge transformations, small
deformations $x^\alpha \to x^\alpha + f^\alpha$ of the coordinates 
employed in the unperturbed spacetime. With regards to transformations
of the spacetime coordinates $x^\alpha$, the horizon quantities 
$\gamma_{AB}$, $\kappa$, $\omega_A$, and ${\cal K}_{AB}$ are a
collection of scalar fields expressed entirely  in terms of the
hypersurface coordinates $(\lambda,\alpha^A)$. As such they are
independent of the spacetime coordinates, and therefore immune
to gauge transformations. As a matter of principle, therefore, all 
horizon quantities are gauge-invariant quantities. The situation,
however, is made more subtle by a matter of practice, our
identification of the generator parameter $\lambda$ with the
advanced-time coordinate $v$ of the underlying spacetime. With this
identification, a transformation of the spacetime coordinates is
necessarily associated with a reparameterization of the null
generators, and the horizon quantities acquire a gauge dependence 
that is inherited from their dependence on reparameterizations. In
this context, the results summarized in the preceding paragraph can be
stated as follows: While the extrinsic geometry of a perturbed event
horizon depends on the choice of gauge, {\it the intrinsic geometry is
gauge-invariant.}   

Our Gauss-Codazzi theory of null hypersurfaces is developed in
Sec.~\ref{sec:diffgeom}. In Sec.~\ref{sec:blackhole} we examine a
nonrotating black hole deformed by an arbitrary distribution of
matter, describe its geometry in terms of a perturbed Schwarzschild
metric, and compute the horizon quantities $\gamma_{AB}$, $\kappa$,
$\omega_A$, and ${\cal K}_{AB}$ to first order in perturbation
theory. In Sec.~\ref{sec:tidal} we specialize our results to tidal
deformations produced by a remote distribution of matter. Adopting a
specific choice of gauge (the ``Killing gauge''), we involve the
vacuum field equations near the horizon to express the horizon
quantities in terms of the well-known master functions $\Psi_{\rm
  even}$ (the Zerilli-Moncrief function) and 
$\Psi_{\rm odd}$ (the Cunningham-Price-Moncrief function) of 
black-hole perturbation theory. In Sec.~\ref{sec:applications} we
consider two applications of our formalism, the first involving a
slowly-varying, quadrupolar tidal field imposed on the black hole, the
second involving a close parabolic encounter between the black hole
and a small orbiting body. An appendix (\ref{app:late-time}) contains
mathematical developments regarding the late-time behaviour of the
horizon quantities.   

\section{Differential geometry of null hypersurfaces} 
\label{sec:diffgeom} 

To guide the development of a theory of perturbed event horizons it is
helpful to formulate a differential geometry of embedded null 
hypersurfaces. The main goal is to arrive at a set of Gauss-Codazzi
equations that apply to the null case instead of being restricted to
usual cases of timelike or spacelike hypersurfaces. The developments
of this section rely on material presented in Secs.~3.1 and 3.11
of Ref.~\cite{poisson:b04}.  

\subsection{Generators, vector basis, and intrinsic coordinates}  
\label{subsec:generator} 

A null hypersurface is generated by a congruence of null 
geodesics that are described by the parametric equations $x^\alpha =
x^\alpha(\lambda,\alpha^A)$, in which $\lambda$ is a running parameter
on each generator, and $\alpha^A = (\alpha,\beta)$ are generator
labels that stay constant on each generator; uppercase latin indices
such as $A$ run from 2 to 3. The null vector field   
\begin{equation} 
k^\alpha = \biggl( \frac{\partial x^\alpha}{\partial \lambda}
\biggr)_{\alpha^A}  
\end{equation}
is tangent to the congruence of null generators, and 
\begin{equation} 
e^\alpha_A = \biggl( \frac{\partial x^\alpha}{\partial \alpha^A}
\biggr)_{\lambda}  
\end{equation}
are spacelike displacements vectors that point from one generator to  
another. These are orthogonal to $k^\alpha$, 
$k_\alpha e^\alpha_A = 0$, and their mutual inner products are 
\begin{equation} 
\gamma_{AB} := g_{\alpha\beta} e^\alpha_A e^\beta_B. 
\label{int_metric} 
\end{equation}  
The definitions imply that the vectors satisfy the Lie-transport
equations 
\begin{equation} 
k^\alpha_{\ ;\beta} e^\beta_A = e^\alpha_{A;\beta} k^\beta, \qquad 
e^\alpha_{A;\beta} e^\beta_B = e^\alpha_{B;\beta} e^\beta_A, 
\label{Lie_transport} 
\end{equation} 
in which a semicolon indicates covariant differentiation in spacetime,
with a connection compatible with $g_{\alpha\beta}$. The basis is
completed with a second null vector $N^\alpha$ that cuts across the
hypersurface; its normalization is chosen so that $N_\alpha k^\alpha 
= -1$, and the vector is also required to satisfy $N_\alpha e^\alpha_A
= 0$.  

We select $(\lambda,\alpha^A)$ as intrinsic coordinates on the
hypersurface. In the spacetime coordinates a displacement within 
the hypersurface is described by $dx^\alpha = k^\alpha\, d\lambda 
+ e^\alpha_A\, d\alpha^A$, and the intrinsic line element is 
\begin{equation} 
ds^2 = \gamma_{AB}\, d\alpha^A d\alpha^B. 
\end{equation} 
This shows that $\gamma_{AB}(\lambda,\alpha^A)$ acts as a metric on
the hypersurface. In this generator-adapted coordinate system, the
degenerate metric is explicitly two-dimensional. We let $\gamma^{AB}$
denote the matrix inverse to $\gamma_{AB}$, and we let $\Gamma^C_{AB}$
be the connection compatible with the two-dimensional metric; the
associated covariant-derivative operator is denoted $\nabla_A$. We use
$\gamma_{AB}$ and its inverse to lower and raise uppercase latin
indices.  

\subsection{Gauss-Weingarten equations} 
\label{subsec:GW} 

The tangent vector fields admit the following set of Gauss-Weingarten
equations: 
\begin{eqnarray} 
k^\alpha_{\ ;\beta} k^\beta &= \kappa\, k^\alpha, 
\label{GWa}\\
k^\alpha_{\ ;\beta} e^\beta_A &= \omega_A k^\alpha 
+ B_A^{\ B} e^\alpha_B = e^\alpha_{A ;\beta} k^\beta, 
\label{GWb} \\ 
e^\alpha_{A;\beta} e^\beta_B &= B_{AB} N^\alpha 
+ {\cal K}_{AB} k^\alpha + \Gamma^C_{AB} e^\alpha_C 
= e^\alpha_{B;\beta} e^\beta_A. 
\label{GWc}
\end{eqnarray}  
These equations define $\kappa$, $\omega_A$, $B_{AB}$, 
${\cal K}_{AB}$, and $\Gamma^C_{AB}$. Explicitly, 
\begin{eqnarray} 
\kappa &= -N_\alpha k^\alpha_{\ ;\beta} k^\beta, \\ 
\omega_A &= -N_\alpha k^\alpha_{\ ;\beta} e^\beta_A, \\ 
B_{AB} &= k_{\alpha;\beta} e^\alpha_A e^\beta_B = B_{BA}, \\ 
{\cal K}_{AB} &= -N_\alpha e^\alpha_{A;\beta} e^\beta_B 
= {\cal K}_{BA}, \\ 
\Gamma_{CAB} &= e_{C \alpha} e^\alpha_{A;\beta} e^\beta_B
= \Gamma_{CBA}, 
\end{eqnarray} 
where $\Gamma_{CAB} := \gamma_{CD} \Gamma^D_{AB}$. These equations
reveal that while $\gamma_{AB}$, $B_{AB}$, and $\Gamma^C_{AB}$
characterize the {\it intrinsic geometry} of the hypersurface,
$\kappa$, $\omega_A$, and ${\cal K}_{AB}$ characterize its 
{\it extrinsic geometry.} 

Each equation in the set (\ref{GWa})--(\ref{GWc}) can be viewed as an
expansion of a vector field (defined on the left-hand-side) in terms
of the hypersurface basis $N^\alpha$, $k^\alpha$, and
$e^\alpha_A$. The absence of terms proportional to $N^\alpha$ in
Eqs.~(\ref{GWa}) and (\ref{GWb}) is a consequence of the fact that
$k^\alpha$ is null everywhere on the hypersurface. The absence of a
term proportional to $e^\alpha_A$ in Eq.~(\ref{GWa}) follows from the
identity $e_{A\alpha} k^\alpha_{\ ;\beta} k^\beta = -k_\alpha
e^\alpha_{A;\beta} k^\beta = 0$, which itself follows from the
orthogonality of $k^\alpha$ and $e^\alpha_A$. Equality of $B_{AB}$ as
defined by Eq.~(\ref{GWb}) and $B_{AB}$ as defined by Eq.~(\ref{GWc})
is confirmed by a similar calculation. Equation (\ref{GWa}) states that
$k^\alpha$ is a geodetic vector field, and $\kappa$ measures the
failure of $\lambda$ to be an affine parameter.   

The definition of Eq.~(\ref{int_metric}) and the Gauss-Weingarten
equations imply that 
\begin{equation} 
\partial_\lambda \gamma_{AB} = 2 B_{AB}. 
\label{gamma_diff} 
\end{equation} 
It is customary to decompose $B_{AB}$ into irreducible components, 
\begin{equation} 
B_{AB} = \frac{1}{2} \Theta \gamma_{AB} + \sigma_{AB}, 
\end{equation} 
with the trace term $\Theta := \gamma^{AB} B_{AB}$
representing the rate of expansion of the congruence of null
generators, and the tracefree term $\sigma_{AB} := B_{AB} -
\frac{1}{2} \Theta \gamma_{AB}$ representing the rate of shear. A
similar decomposition could also be introduced for ${\cal K}_{AB}$.   

The Gauss-Weingarten equations also imply that 
\begin{eqnarray} 
N^\alpha_{\ ;\beta} k^\beta &= -\kappa N^\alpha 
+ \omega^A e^\alpha_A, \\ 
N^\alpha_{\ ;\beta} e^\beta_A &= -\omega_A N^\alpha 
+ {\cal K}_A^{\ B} e^\alpha_B. 
\end{eqnarray} 
These equations govern the behaviour of the transverse vector on the
hypersurface.  

\subsection{Gauss-Codazzi equations} 
\label{subsec:GC} 
 
It is straightforward, following the methods described in Sec.~3.5 of
Ref.~\cite{poisson:04b}, to derive from Eqs.~(\ref{GWa})--(\ref{GWc})
a set of Gauss-Codazzi equations which express projections of the
spacetime Riemann tensor in terms of geometric quantities defined on
the null hypersurface. We have 
\begin{eqnarray} 
\fl
R_{\mu\nu\lambda\alpha} k^\mu N^\nu k^\lambda e^\alpha_A 
&= \partial_\lambda \omega_A - \partial_A \kappa + B_A^{\ B} \omega_B,
\label{GCa} \\ 
\fl
R_{\mu\nu\alpha\beta} k^\mu N^\nu e^\alpha_A e^\beta_B 
&= \nabla_A \omega_B - \nabla_B \omega_A - B_A^{\ C} {\cal K}_{CB}
+ B_B^{\ C} {\cal K}_{CA}, 
\label{GCb} \\ 
\fl
R_{\mu\alpha\nu\beta} k^\mu e^\alpha_A N^\nu e^\beta_B 
&= -\partial_\lambda {\cal K}_{AB} - \kappa {\cal K}_{AB}  
+ \nabla_A \omega_B + \omega_A \omega_B 
+ {\cal K}_A^{\ C} B_{CB}, 
\label{GCc} \\
\fl
R_{\mu\alpha\nu\beta} k^\mu e^\alpha_A k^\nu e^\beta_B 
&= -\partial_\lambda B_{AB} + \kappa B_{AB} + B_A^{\ C} B_{CB}, 
\label{GCd} \\
\fl
R_{\mu\alpha\beta\gamma} k^\mu e^\alpha_A e^\beta_B e^\gamma_C 
&= \nabla_C B_{AB} - \nabla_B B_{AC} - \omega_C B_{AB} + \omega_B B_{AC}, 
\label{GCe} \\ 
\fl
R_{\mu\alpha\beta\gamma} N^\mu e^\alpha_A e^\beta_B e^\gamma_C
&= \nabla_C {\cal K}_{AB} - \nabla_B {\cal K}_{AC} + \omega_C {\cal K}_{AB} 
- \omega_B {\cal K}_{AC}, 
\label{GCf} \\ 
\fl
R_{\alpha\beta\gamma\delta} e^\alpha_A e^\beta_B e^\gamma_C e^\delta_D 
&= \frac{1}{2} {\cal R} \bigl( \gamma_{AC} \gamma_{BD} 
- \gamma_{AD} \gamma_{BC} \bigr)+ B_{AC} {\cal K}_{BD} 
- B_{AD} {\cal K}_{BC} 
\nonumber \\ \fl & \quad \mbox{}  
+ {\cal K}_{AC} B_{BD} - {\cal K}_{AD} B_{BC}, 
\label{GCg} 
\end{eqnarray} 
where ${\cal R}$ is the Ricci scalar associated with the
two-dimensional metric $\gamma_{AB}$. To arrive at these equations we
used the fact that the Riemann tensor on a two-dimensional metric
space can always be expressed as ${\cal R}_{ABCD} =  \frac{1}{2} 
{\cal R} ( \gamma_{AC} \gamma_{BD} - \gamma_{AD} \gamma_{BC})$. We
also relied on the identity $\gamma_{CD} \partial_\lambda \Gamma^D_{AB} 
= \nabla_A B_{BC} + \nabla_B B_{AC} - \nabla_C B_{AB}$, which can be derived on the
basis of Eq.~(\ref{gamma_diff}).  

Insertion of the Gauss-Codazzi equations within the identity 
\begin{equation} 
R_{\mu\nu} = g^{\alpha\beta} R_{\alpha\mu\beta\nu} 
= \bigl( -k^\alpha N^\beta - N^\alpha k^\beta + \gamma^{AB}
e^\alpha_A e^\beta_B \bigr) R_{\alpha\mu\beta\nu} 
\end{equation} 
produces the following components of the Ricci tensor: 
\begin{eqnarray} 
\fl
R_{\mu\nu} k^\mu k^\nu &= -\partial_\lambda \Theta 
+ \kappa \Theta - B_{AB} B^{AB}, 
\label{GCRa} \\ 
\fl
R_{\mu\alpha} k^\mu e^\alpha_A &= \partial_\lambda \omega_A 
- \partial_A \kappa - \partial_A \Theta + \nabla_B B_A^{\ B} 
+ \Theta \omega_A, 
\label{GCRb} \\ 
\fl
R_{\alpha\beta} e^\alpha_A e^\beta_B &=
2 \bigl( \partial_\lambda {\cal K}_{AB} + \kappa {\cal K}_{AB} \bigr) 
- \bigl( \nabla_A \omega_B + \nabla_B \omega_A \bigr) - 2 \omega_A \omega_B 
\nonumber \\ \fl & \quad \mbox{}
- 2 \bigl( {\cal K}_A^{\ C} B_{CB} + {\cal K}_B^{\ C} B_{CA} \bigr) 
+ \Theta {\cal K}_{AB} + {\cal K} B_{AB} 
+ \frac{1}{2} {\cal R} \gamma_{AB}, 
\label{GCRc} 
\end{eqnarray} 
where ${\cal K} := \gamma^{AB} {\cal K}_{AB}$. 

By involving the Einstein field equations, Eq.~(\ref{GCRa}) can be
turned into Raychaudhuri's equation, 
\begin{equation} 
\partial_\lambda \Theta = \kappa \Theta - \frac{1}{2} \Theta^2 
- \sigma_{AB} \sigma^{AB} - 8\pi T_{\alpha\beta} k^\alpha k^\beta, 
\label{ray1} 
\end{equation} 
with $T_{\alpha\beta} k^\alpha k^\beta$ representing the flux of
matter across the null hypersurface. And by extracting the tracefree
piece of Eq.~(\ref{GCd}) we obtain an analogous equation for the shear 
tensor, 
\begin{equation} 
\partial_\lambda \sigma^A_{\ B} = (\kappa-\Theta) \sigma^A_{\ B} 
- C^A_{\ B},
\label{ray2}
\end{equation} 
where $C_{AB} := 
C_{\mu\alpha\nu\beta} k^\mu e^\alpha_{A} k^\nu e^\beta_{B}$ are
components of the Weyl tensor. 

\subsection{Reparameterizations} 
\label{subsec:reparam} 

The geometric quantities $\gamma_{AB}$, $B_{AB}$, $\kappa$,
$\omega_A$, and ${\cal K}_{AB}$ all refer to a selected
parameterization $(\lambda,\alpha^A)$ of the null generators. We first
examine how these quantities change under a reparameterization of the
form 
\begin{equation} 
\lambda \to \bar{\lambda}(\lambda,\alpha^A), 
\end{equation} 
which represents an independent change of parameter on each
generator. The differential form of the transformation is expressed as 
\begin{equation} 
d\bar{\lambda} = e^{-\beta} \bigl( d\lambda - c_A\, d\alpha^A \bigr), 
\end{equation} 
with 
\begin{equation} 
e^{-\beta} := \biggl( \frac{\partial \bar{\lambda}}{\partial \lambda}
\biggr)_{\alpha^A},\qquad 
- e^{-\beta} c_A := \biggl( \frac{\partial \bar{\lambda}}
{\partial \alpha^A} \biggr)_{\lambda}. 
\label{partial_dervs} 
\end{equation} 
These are functions of $(\lambda,\alpha^A)$ on the hypersurface, and
the notation was chosen so as to simplify our expressions below. The
inverse transformation is $d\lambda = e^\beta d\bar{\lambda} 
+ c_A\, d\alpha^A$. 

As we saw previously, a displacement on the hypersurface is described
by $dx^\alpha = k^\alpha\, d\lambda + e^\alpha_A\, d\alpha^A$, but the
reparameterization brings this to the new form $dx^\alpha 
= \bar{k}^\alpha\, d\bar{\lambda} + \bar{e}^\alpha_A\, d\alpha^A$,
with 
\begin{equation} 
\bar{k}^\alpha = e^\beta k^\alpha, \qquad 
\bar{e}^\alpha_A = e^\alpha_A + c_A k^\alpha. 
\end{equation} 
These vectors have the same interpretation as the old vectors:
$\bar{k}^\alpha$ is still tangent to the congruence of null
generators, but is renormalized so as to reflect the new
parameterization, and $\bar{e}^\alpha_A$ still points from generator
to generator. It is easy to show that the new transverse vector must
be given by  
\begin{equation} 
\bar{N}^\alpha = e^{-\beta} \bigl( N^\alpha 
+ \sfrac{1}{2} c^A c_A k^\alpha  + c^A e^\alpha_A \bigr) 
\end{equation} 
to satisfy its defining relations. The inverse transformations are
$k^\alpha = e^{-\beta} \bar{k}^\alpha$, $e^\alpha_A = \bar{e}^\alpha_A
- e^{-\beta} c_A \bar{k}^\alpha$, and 
$N^\alpha = e^\beta \bar{N}^\alpha + \frac{1}{2} e^{-\beta} c^A c_A
\bar{k}^\alpha - c^A \bar{e}^\alpha_A$.  

The reparameterization produces the following changes in the geometric
quantities: 
\begin{eqnarray} 
\bar{\gamma}_{AB} &= \gamma_{AB}, 
\label{repara_a} \\ 
\bar{B}_{AB} &= e^\beta B_{AB}, 
\label{repara_b} \\ 
\bar{\kappa} &= e^\beta (\kappa + \partial_\lambda \beta ), 
\label{repara_c} \\  
\bar{\omega}_A &= \omega_A - B_A^{\ B} c_B + \kappa c_A 
+ c_A \partial_\lambda \beta + \partial_A \beta,
\label{repara_d} \\ 
\bar{\cal K}_{AB} &= e^{-\beta} \bigl( {\cal K}_{AB} 
+ \omega_A c_B + \omega_B c_A + \kappa c_A c_B
+ c_B \partial_\lambda c_A + \nabla_B c_A 
\nonumber \\  & \quad \mbox{} 
+ \sfrac{1}{2} c^C c_C B_{AB} - B_A^{\ C} c_C c_B 
- B_B^{\ C} c_C c_A \bigr).  
\label{repara_e}
\end{eqnarray} 
The term $c_B \partial_\lambda c_A  + \nabla_B c_A$ in the last equation is
not manifestly symmetric in the pair of indices $AB$. With the
definitions of Eqs.~(\ref{partial_dervs}), however, we find that this
can be expressed in the form 
\begin{eqnarray}
\fl 
c_B \partial_\lambda c_A  + \nabla_B c_A 
&= -e^\beta \frac{\partial^2 \bar{\lambda}}
  {\partial \alpha^A \partial \alpha^B}
+ e^{2\beta} \frac{\partial^2 \bar{\lambda}}
  {\partial \lambda \partial \alpha^A} 
  \frac{\partial \bar{\lambda}}{\partial \alpha^B} 
+ e^{2\beta} \frac{\partial^2 \bar{\lambda}}
  {\partial \lambda \partial \alpha^B} 
  \frac{\partial \bar{\lambda}}{\partial \alpha^A} 
\nonumber \\ \fl & \quad \mbox{}
- e^{3\beta} \frac{\partial^2 \bar{\lambda}}
  {\partial \lambda^2} 
  \frac{\partial \bar{\lambda}}{\partial \alpha^A} 
  \frac{\partial \bar{\lambda}}{\partial \alpha^B} 
- \Gamma^C_{AB} c_C, 
\end{eqnarray} 
which reveals the required symmetry. An additional change produced by
the reparameterization is $\bar{\Gamma}^C_{AB} = \Gamma^C_{AB} 
+ B_A^{\ C} c_B + B_B^{\ C} c_A - B_{AB} c^C$.  

In the case of infinitesimal transformations described by
$\bar{\lambda} = \lambda + \delta \lambda(\lambda,\alpha^A)$, the
partial derivatives are captured by $\delta \beta := -\partial_\lambda
\delta \lambda$ and $\delta c_A := -\partial_A \delta \lambda$, and
the transformations of Eqs.~(\ref{repara_a})--(\ref{repara_e})
simplify. For the purposes of an application of the formalism
presented below, we assume that the geometric quantities can be
expressed as  
\begin{eqnarray} 
\gamma_{AB} &= \gamma^0_{AB} + \delta \gamma_{AB}, \\
B_{AB} &= \delta B_{AB}, \\ 
\kappa &= \kappa_0 + \delta \kappa, \\ 
\omega_A &= \delta \omega_A, \\ 
{\cal K}_{AB} &= {\cal K}^0_{AB} + \delta {\cal K}_{AB}, 
\end{eqnarray} 
where the ``background quantities'' $\gamma^0_{AB}$, $\kappa_0$, and
${\cal K}^0_{AB}$ are assumed to be $\lambda$-independent, 
and where $\delta \gamma_{AB}$, $\delta B_{AB}$, 
$\delta \kappa$, $\delta \omega_A$, and $\delta {\cal K}_{AB}$ are
$\lambda$-dependent ``perturbations.'' In this restricted context the
transformations reduce to 
\begin{eqnarray} 
\delta\bar{\gamma}_{AB} &= \delta \gamma_{AB}, 
\label{repara_inf_a} \\ 
\delta\bar{B}_{AB} &= \delta B_{AB}, 
\label{repara_inf_b} \\ 
\delta\bar{\kappa} &= \delta \kappa 
+ \partial_\lambda \delta\beta + \kappa_0 \delta\beta, 
\label{repara_inf_c} \\ 
\delta\bar{\omega}_A &= \delta\omega_A 
+ \partial_A \delta\beta + \kappa_0 \delta c_A, 
\label{repara_inf_d} \\ 
\delta\bar{\cal K}_{AB} &= \delta{\cal K}_{AB} 
- {\cal K}^0_{AB} \delta \beta 
+ \nabla_B \delta c_A.
\label{repara_inf_e}  
\end{eqnarray} 
In the last equation the covariant derivative $\nabla_B$ is evaluated with
a connection compatible with the background metric $\gamma^0_{AB}$. 

We next examine the possibility of transforming the generator
labels. A general transformation of the form $\alpha^A \to
\bar{\alpha}^A(\lambda,\alpha^B)$ is excluded,
because the dependence upon $\lambda$ would imply that
$\bar{\alpha}^A$ is not constant on each generator, in violation of
its defining property. The remaining freedom is a rigid transformation 
of the form $\alpha^A \to \bar{\alpha}^A(\alpha^B)$, upon which
scalars such as $\kappa$ remain invariant, while tensors such as
$\omega_A$ and $\gamma_{AB}$ transform in the usual way. In 
particular, for infinitesimal transformations of the form
$\bar{\alpha}^A = \alpha^A + \delta \alpha^A$, the metric tensor
transforms as 
\begin{equation} 
\bar{\gamma}_{AB}(\bar{\alpha}^C) = \gamma_{AB}(\bar{\alpha}^C) 
- \nabla_A \delta\alpha_B - \nabla_B \delta\alpha_A, 
\label{relabeling} 
\end{equation} 
where $\nabla_A$ refers to $\gamma_{AB}$, and $\delta\alpha_A =
\gamma_{AB} \delta\alpha^B$. In this formulation the original metric
is expressed as a function of the new coordinates (instead of the
original coordinates), and the transformation takes the standard
appearance of a gauge transformation.  

\section{Deformed black hole} 
\label{sec:blackhole} 

We consider a nonrotating black hole perturbed by a distribution of
matter. The perturbation is sufficiently small that we can describe it  
within linearized perturbation theory, and to achieve this we rely on
the formulation of the theory provided in
Ref.~\cite{martel-poisson:05}. The matter is 
either flowing across the event horizon, in which case the
perturbation is sourced by matter, or it is situated outside the black
hole's immediate neighborhood, in which case the perturbation is in
vacuum and describes a tidal deformation of the black hole.   

\subsection{Spacetime metric} 
\label{subsec:metric} 

The metric of the unperturbed spacetime is Schwarzschild's solution
expressed in Eddington-Finkelstein coordinates, 
\begin{equation} 
g^0_{\alpha\beta}\, dx^\alpha dx^\beta = -f\, dv^2 + 2\, dvdr  
+ r^2 \bigl( d\theta^2 + \sin^2\theta\, d\phi^2 \bigr), 
\end{equation} 
with $f := 1-2M/r$. We let $x^a = (v,r)$ and $\theta^A =
(\theta,\phi)$. The metric on the unit two-sphere is $\Omega_{AB}\,
d\theta^A d\theta^B = d\theta^2 + \sin^2\theta\, d\phi^2$, and its
inverse is denoted $\Omega^{AB}$; covariant differentiation compatible
with $\Omega_{AB}$ is denoted $D_A$. 

The metric perturbation is denoted $p_{\alpha\beta}$, and it is
decomposed in tensorial spherical harmonics (as defined in
Ref.~\cite{martel-poisson:05}). In the even-parity sector we have  
\begin{eqnarray} 
p_{ab} &= h_{ab}(v,r) Y(\theta^A), 
\label{even_a} \\ 
p_{aB} &= j_a(v,r) Y_B(\theta^A), 
\label{even_b} \\ 
p_{AB} &= r^2 K(v,r) \Omega_{AB} Y(\theta^A) 
+ r^2 G(v,r) Y_{AB}(\theta^A),  
\label{even_c} 
\end{eqnarray} 
with $Y(\theta^A)$ denoting standard spherical-harmonic functions,  
$Y_A := D_A Y$, and $Y_{AB} := [D_A D_B + \frac{1}{2} \ell(\ell+1)
\Omega_{AB}] Y$. In the odd-parity sector we have 
\begin{eqnarray} 
p_{ab} &= 0, 
\label{odd_a} \\
p_{aB} &= h_a(v,r) X_B(\theta^A), 
\label{odd_b} \\  
p_{AB} &= h_2(v,r) X_{AB}(\theta^A), 
\label{odd_c} 
\end{eqnarray} 
where $X_A := -\varepsilon_A^{\ B} D_B Y$ and $X_{AB} :=
\frac{1}{2}(D_A X_B + D_B X_A)$, with $\varepsilon_{AB}$ denoting the
Levi-Civita tensor (with component $\varepsilon_{\theta\phi} =
\sin\theta$) on the unit two-sphere. The tensorial harmonics $Y_{AB}$
and $X_{AB}$ are both symmetric and tracefree. The spherical-harmonic
labels $\ell m$ are suppressed, and so is summation over these labels.
The complete metric of the perturbed spacetime is 
$g_{\alpha\beta} = g^0_{\alpha\beta} + p_{\alpha\beta}$.  

Under an even-parity gauge transformation generated by the vector
field $f_a = \eta_a(v,r) Y$ and $f_A = r^2 \eta^{\rm even}(v,r) Y_A$,
the perturbation fields change according to   
\begin{eqnarray} 
\Delta h_{vv} &= -2 \partial_v \eta_v 
+ \frac{2M}{r^2} \eta_v + \frac{2Mf}{r^2} \eta_r, \\ 
\Delta h_{vr} &= -\partial_r \eta_v 
- \partial_v \eta_r - \frac{2M}{r^2} \eta_r, \\
\Delta h_{rr} &= -2 \partial_r \eta_r, \\
\Delta j_v &= -r^2 \partial_v \eta^{\rm even} - \eta_v, \\ 
\Delta j_r &= -r^2 \partial_r \eta^{\rm even} - \eta_r, \\
\Delta K &= -\frac{2f}{r} \eta_r - \frac{2}{r} \eta_v 
+ \ell(\ell+1) \eta^{\rm even}, \\ 
\Delta G &= -2 \eta^{\rm even}. 
\end{eqnarray} 
Under an odd-parity gauge transformation generated by the vector 
field $f_a = 0$ and $f_A = r^2 \eta^{\rm odd}(v,r) X_A$, the
perturbation fields change according to  
\begin{eqnarray} 
\Delta h_{v} &= -r^2 \partial_v \eta^{\rm odd}, \\
\Delta h_{r} &= - r^2 \partial_r \eta^{\rm odd}, \\
\Delta h_2 &= - 2r^2 \eta^{\rm odd}. 
\end{eqnarray}  
These transformations will play a role in the forthcoming
developments. 

\subsection{Deformed horizon}
\label{subsec:horizon}  

The description of the deformed horizon relies on the geometrical
methods reviewed in Sec.~\ref{sec:diffgeom}. The event
horizon is traced by its null generators, which are identified by
constant labels $\alpha^A = (\alpha,\beta)$; we use 
$\lambda \equiv v$ as a running parameter on each generator, and
$(v,\alpha^A)$ forms a system of intrinsic coordinates on the
horizon. The parametric equations that describe the horizon's position
in the unperturbed spacetime are $v = v$, $r = 2M$, and 
$\theta^A = \alpha^A$. In the perturbed spacetime we have instead    
\begin{equation} 
\fl
v = v, \qquad
r = 2M\bigl[1 + B(v,\alpha^A) \bigr], \qquad 
\theta^A = \alpha^A + \Xi^A(v,\alpha^A), 
\label{hor_param} 
\end{equation} 
where $2M B$ and $\Xi^A$ are the components of a Lagrangian
displacement vector. This vector takes the horizon point identified by 
$(v,\alpha^A)$ in the original spacetime to a point also identified by
$(v,\alpha^A)$ in the perturbed spacetime. We express the displacement
fields as  
\begin{equation} 
\fl
B = b(v) Y(\alpha^A), \qquad 
\Xi^A =\Omega^{AB} \bigl[ \xi^{\rm even}(v) Y_B(\alpha^A) 
+ \xi^{\rm odd}(v) X_B(\alpha^A) \bigr], 
\label{B_Xi} 
\end{equation} 
in which $\Omega^{AB}$ is expressed in terms of the intrinsic
coordinates $\alpha^A$. As previously we suppress the $\ell m$ labels,
as well as summation over these labels.  

The parametric equations (\ref{hor_param}) allow us to calculate the 
basis vectors 
\begin{equation} 
k^\alpha := \biggl( \frac{\partial x^\alpha}{\partial v}
\biggr)_{\alpha^A}, \qquad 
e^\alpha_{(A)} := \biggl( \frac{\partial x^\alpha}{\partial \alpha^A}
\biggr)_{v}.  
\end{equation} 
In this section we place brackets around a basis index (which refers 
to the intrinsic coordinates $\alpha^A$) to distinguish it from a
coordinate index (which refers to the spacetime coordinates
$\theta^A$).  Explicitly, 
\begin{eqnarray} 
k^v &= 1, \\ 
k^r &= 2M \partial_v B = 2M \dot{b}\, Y, \\ 
k^A &= \partial_v \Xi^A = \Omega^{AB} \bigl[ \dot{\xi}^{\rm even} Y_B  
+ \dot{\xi}^{\rm odd} X_B \bigr],  
\end{eqnarray} 
in which an overdot indicates differentiation with respect to $v$, and 
\begin{equation} 
e^v_{(A)} = 0, \qquad 
e^r_{(A)} = 2M \partial_A B, \qquad 
e^A_{(B)} = \delta^A_{\ B} + \partial_B \Xi^A. 
\label{e_vectors} 
\end{equation} 
The null condition $k_\alpha k^\alpha = 0$ gives rise to the first
horizon equation,  
\begin{equation} 
b - 4M \dot{b} = h_{vv}(v,2M), 
\label{horizon_equation} 
\end{equation} 
and the conditions $k_\alpha e^\alpha_{(A)} = 0$ give rise to a second
set of horizon equations,  
\begin{eqnarray} 
\dot{\xi}^{\rm even} &= -(2M)^{-2}\bigl[ j_v(v,2M) + 2Mb(v) \bigr], 
\label{hor_eq_a} \\   
\dot{\xi}^{\rm odd} &= -(2M)^{-2} h_v(v,2M).
\label{hor_eq_b}  
\end{eqnarray}  
These equations, along with appropriate choices of boundary
conditions, fully determine the description of the deformed horizon. 

The basis can be completed with a transverse vector $N^\alpha$ that
satisfies the relations $N_\alpha N^\alpha = 0$, 
$N_\alpha k^\alpha = -1$, and $N_\alpha e^\alpha_{(A)} = 0$. A simple
computation reveals that the components of this vector are given by  
\begin{eqnarray} 
N^v &= \frac{1}{2} h_{rr} Y, \\ 
N^r &= -1 + h_{vr} Y, \\ 
N^A &= (2M)^{-2} \Omega^{AB} \bigl( j_r Y_B + h_r X_B \bigr), 
\label{N_vector} 
\end{eqnarray} 
where all perturbation fields are evaluated at $r=2M$. The covariant
components of the vector are $N_v = -1$, $N_r = -\frac{1}{2} h_{rr}
Y$, and $N_A = 0$. 

To identify the correct solutions to the horizon equations we
imagine first an artificial situation in which the perturbation is
switched off at times larger than $v_1$. The spacetime for 
$v > v_1$ is described by the Schwarzschild metric, and for these
times the event horizon is correctly identified with the hypersurface
$r=2M$. To locate the event horizon at times $v < v_1$ we must
smoothly extend $r=2M$ backwards in time, to a null hypersurface in
the perturbed spacetime. This surface is described by 
Eq.~(\ref{hor_param}), with $b(v)$ restricted to vanish for
$v>v_1$. The appropriate solution to Eq.~(\ref{horizon_equation}) is
therefore  
\begin{equation} 
b(v) = \kappa_0 \int_v^\infty e^{-\kappa_0 (v'-v)} 
h_{vv}(v',2M)\, dv',   
\label{teleo_solution1} 
\end{equation} 
where $\kappa_0 := (4M)^{-1}$ is the surface gravity of the
unperturbed black hole. The upper limit of integration was extended to
$v=\infty$ because, by the stated assumptions on the perturbation,
$h_{vv}$ is zero in the interval $v_1 < v' < \infty$. At this stage,
however, the artifice can be removed and Eq.~(\ref{teleo_solution1})
be adopted as the appropriate solution to Eq.~(\ref{horizon_equation})
even when the perturbation does not switch off at $v=v_1$. The
perturbation must still fall off sufficiently fast that the integral
converges, and under these conditions $b(v)$ will approach zero as 
$v \to \infty$. Because Eq.~(\ref{teleo_solution1}) reflects a choice
of final condition, it is known as a {\it teleological solution} to
the horizon equation.  

The teleological solutions to Eqs.~(\ref{hor_eq_a}) and
(\ref{hor_eq_b}) are 
\begin{eqnarray} 
\xi^{\rm even}(v) &= (2M)^{-2} \int_v^\infty 
\bigl[ j_v(v',2M) + 2Mb(v') \bigr]\, dv', 
\label{teleo_sol_a} \\ 
\xi^{\rm odd}(v) &= (2M)^{-2} \int_v^\infty h_v(v',2M) \, dv'. 
\label{teleo_sol_b}
\end{eqnarray}  
The behaviour of the horizon generators in the perturbed spacetime is
now completely determined. The solutions to the horizon equations
imply that in general, the event horizon leads the perturbation by a
time interval of order $\kappa_0^{-1} = 4M$. 

\subsection{Horizon's intrinsic geometry} 
\label{subsec:intrinsic} 

As described in Sec.~\ref{sec:diffgeom}, the intrinsic geometry of the
event horizon is characterized by the induced metric $\gamma_{AB}$,
which is expressed in the intrinsic coordinates $(v,\alpha^A)$
attached to the null generators. According to Eq.~(\ref{gamma_diff}), 
the $v$-derivative of the induced metric satisfies 
\begin{equation} 
\partial_v \gamma_{AB} = 2 B_{AB} = \Theta \gamma_{AB} 
+ 2 \sigma_{AB}, 
\label{exp_shear} 
\end{equation} 
and this equation defines the expansion scalar $\Theta$ and shear 
tensor $\sigma_{AB}$ associated with the congruence of null
generators. The expansion, in particular, can be computed as 
$\Theta = \frac{1}{2} \gamma^{-1} \partial_v \gamma$, where  
$\gamma := \mbox{det}[\gamma_{AB}]$. 

A computation of the horizon metric involves the substitution of
Eq.~(\ref{e_vectors}) into Eq.~(\ref{int_metric}). The computation
must account for the fact that while the spacetime metric is expressed
in terms of the coordinates $(v,r,\theta^A)$, the horizon metric will
be expressed in terms of the intrinsic coordinates $(v,\alpha^A)$. A
piece of the computation that requires some care involves
$\Omega_{AB}(\theta^A)$, which must be written as
$\Omega_{AB}(\alpha^A + \Xi^A) = \Omega_{AB} +
\Xi^C \partial_C \Omega_{AB}$, with the right-hand side expressed
in terms of $\alpha^A$. With this accounted for, we find that the
horizon metric is  
\begin{equation} 
\fl
\gamma_{AB} = (2M)^2 \Omega_{AB} + (2M)^2 \bigl( 2 B \Omega_{AB} 
+ \Omega_{BC} D_A \Xi^C + \Omega_{AC} D_B \Xi^C \bigr) + p_{AB}. 
\end{equation} 
With Eqs.~(\ref{even_a})--(\ref{odd_c}) and (\ref{B_Xi}), this becomes  
\begin{equation} 
\gamma_{AB} = (2M)^2 \bigl( \Omega_{AB} 
+ \gamma^{\rm trace}\Omega_{AB} Y 
+ \gamma^{\rm even} Y_{AB} 
+ \gamma^{\rm odd} X_{AB} \bigr), 
\label{hor_metric} 
\end{equation} 
where 
\begin{eqnarray} 
\gamma^{\rm trace} &:= 2b(v) - \ell(\ell+1) \xi^{\rm even}(v) 
+ K(v,2M), \\ 
\gamma^{\rm even} &:= 2\xi^{\rm even}(v) + G(v,2M), \\ 
\gamma^{\rm odd} &:= 2 \xi^{\rm odd}(v) + (2M)^{-2} h_2(v,2M).  
\end{eqnarray} 
The square root of the metric determinant is given by
$\sqrt{\gamma} = (2M)^2 \sin\alpha( 1 + \gamma^{\rm trace} Y )$, 
with $\alpha$ denoting the intrinsic polar angle on the horizon.  

It follows from these equations that the expansion scalar is 
\begin{equation} 
\Theta = \dot{\gamma}^{\rm trace} Y, 
\label{expansion} 
\end{equation} 
while the shear tensor is 
\begin{equation} 
\sigma_{AB} = \frac{1}{2} (2M)^2 \bigl( \dot{\gamma}^{\rm even} Y_{AB} 
+ \dot{\gamma}^{\rm odd} X_{AB} \bigr). 
\label{shear}
\end{equation} 
The expressions for $\dot{\gamma}^{\rm trace}$, 
$\dot{\gamma}^{\rm even}$, and $\dot{\gamma}^{\rm odd}$ can be 
simplified with the help of
Eqs.~(\ref{horizon_equation})--(\ref{hor_eq_b}). We obtain 
\begin{eqnarray} 
\fl
\dot{\gamma}^{\rm trace} &= 
(2M)^{-1} \Bigl\{ [\ell(\ell+1) +1] b(v) 
- h_{vv}(v,2M) + \ell(\ell+1) (2M)^{-1} j_v(v,2M) 
\nonumber \\ \fl & \quad \mbox{}
+ (2M) \partial_v K(v,2M) \Bigr\}, \\ 
\fl
\dot{\gamma}^{\rm even} &= 
(2M)^{-1} \Bigl\{ -2b(v) - 2(2M)^{-1} j_v(v,2M) 
+ 2M\partial_v G(v,2M)  \Bigr\}, \\ 
\fl
\dot{\gamma}^{\rm odd} &= 
(2M)^{-1} \Bigl\{ -2 (2M)^{-1} h_v(v,2M) 
+ (2M)^{-1} \partial_v h_2(v,2M) \Bigr\}. 
\end{eqnarray} 

The Ricci curvature associated with the metric of
Eq.~(\ref{hor_metric}) is given by 
\begin{equation} 
\fl
{\cal R} = \frac{1}{2M^2} \biggl\{ 1 + \frac{1}{2} (\ell-1)(\ell+2) 
\Bigl[ \gamma^{\rm trace}
+ \frac{1}{2} \ell(\ell+1) \gamma^{\rm even} \Bigr] Y(\alpha^A)
\biggl\}.  
\label{ricci} 
\end{equation} 
This indicates that the metric's geometrical information is contained
within $\gamma^{\rm trace} + \frac{1}{2} \ell(\ell+1) 
\gamma^{\rm even} = 2b + K + \frac{1}{2} \ell(\ell+1) G$; the
remaining information is entirely about the choice of intrinsic
coordinates, in particular, the fact that they are attached to the
horizon's null generators. To flesh out this last point we recall
that, according to the discussion near the end of 
Sec.~\ref{subsec:reparam}, the freedom
to change the generator labels $\alpha^A$ is limited to a
$v$-independent rotation of the form 
$\alpha^A \to \bar{\alpha}^A(\alpha^B)$. For
infinitesimal changes $\bar{\alpha}^A = \alpha^A + \delta \alpha^A$ 
the transformation is described by Eq.~(\ref{relabeling}). If we choose 
\begin{equation} 
\delta \alpha^A = \Omega^{AB} \bigl( \zeta^{\rm even} Y_B 
+ \zeta^{\rm odd} X_B \bigr), 
\end{equation} 
where $\zeta^{\rm even}$ and $\zeta^{\rm odd}$ are constants, 
we find that $\gamma^{\rm trace}$, $\gamma^{\rm even}$, and 
$\gamma^{\rm odd}$ change according to 
\begin{eqnarray}
\gamma^{\rm trace} &\to \gamma^{\rm trace} 
+ \ell(\ell+1) \zeta^{\rm even}, \\  
\gamma^{\rm even} &\to \gamma^{\rm even} 
- 2\zeta^{\rm even}, \\  
\gamma^{\rm odd} &\to \gamma^{\rm odd} 
- 2\zeta^{\rm even}. 
\end{eqnarray}   
We observe that the combination $\gamma^{\rm trace} 
+ \frac{1}{2} \ell(\ell+1) \gamma^{\rm even}$ is unaffected by the
transformation, which confirms its role as carrier of geometric
information. At any given time (but at only one such time),
$\zeta^{\rm even}$ and $\zeta^{\rm odd}$ can be chosen so as as to
make $\gamma^{\rm even}$ and $\gamma^{\rm odd}$ vanish. At this time,
say $v = v_0$, we have that the horizon metric is given by 
\begin{equation} 
\gamma_{AB}(v_0,\alpha^A) = (2M)^2 \Omega_{AB} 
\bigl[ 1 + \gamma^{\rm trace} Y(\alpha^A) \bigr], 
\end{equation} 
with $\gamma^{\rm trace} := \gamma^{\rm trace}_{\rm new} 
= \gamma^{\rm trace}_{\rm old} + \frac{1}{2} \ell(\ell+1) 
\gamma^{\rm even}_{\rm old}$, or 
\begin{equation} 
\gamma^{\rm trace}(v_0) = 2b(v_0) + K(v_0,2M) 
+ \frac{1}{2} \ell(\ell+1) G(v_0,2M). 
\end{equation} 
At other times $v \neq v_0$, the tracefree terms proportional to
$Y_{AB}$ and $X_{AB}$ will no longer vanish, and the metric will
return to its general form of Eq.~(\ref{hor_metric}).    

\subsection{Horizon's extrinsic geometry} 
\label{subsec:extrinsic} 

The horizon's extrinsic geometry is characterized by $\kappa$,
$\omega_A$, and ${\cal K}_{AB}$, as defined by
Eqs.~(\ref{GWa})--(\ref{GWc}). Computation reveals that  
\begin{equation} 
\kappa = \kappa_0 \bigl( 1 + k Y \bigr),  
\end{equation} 
where $\kappa_0 = (4M)^{-1}$ is the unperturbed surface gravity, and  
\begin{equation} 
k = -\bigl( 2M \partial_r h_{vv} 
- 4M \partial_v h_{vr} + h_{vr} + 2b \bigr). 
\end{equation} 
We also get  
\begin{equation} 
\omega_A = \omega^{\rm even} Y_A + \omega^{\rm odd} X_A, 
\end{equation}
with 
\begin{eqnarray}
\omega^{\rm even} &= \frac{1}{2} h_{vr} - \frac{1}{2} \partial_r j_v 
+ (2M)^{-1} j_v + \frac{1}{2} \partial_v j_r + b, \\
\omega^{\rm odd} &= -\frac{1}{2} \partial_r h_v + (2M)^{-1} h_v 
+ \frac{1}{2} \partial_v h_r. 
\end{eqnarray}
And finally, we get 
\begin{equation} 
K_{AB} = -2M \Omega_{AB} + {\cal K}^{\rm trace} \Omega_{AB} Y 
+ {\cal K}^{\rm even} Y_{AB} + {\cal K}^{\rm odd} X_{AB}, 
\end{equation} 
with 
\begin{eqnarray} 
{\cal K}^{\rm trace} &=  \ell(\ell+1) 2M \xi^{\rm even} + 2M h_{vr} 
- \frac{1}{2} \ell(\ell+1) j_r 
\nonumber \\ & \quad \mbox{}
- \frac{1}{2} (2M)^2 \partial_r K 
- 2M K - 2M b, \\ 
{\cal K}^{\rm even} &= -4M \xi^{\rm even} + j_r 
- \frac{1}{2} (2M)^2 \partial_r G - 2M G, \\ 
{\cal K}^{\rm odd} &= -4M \xi^{\rm odd} + h_r 
- \frac{1}{2} \partial_r h_2.
\end{eqnarray} 
In these expressions, all perturbation fields and their derivatives
are evaluated at $r=2M$, and all spherical harmonics are expressed as
functions of $\alpha^A$. We note that the computation of 
${\cal K}_{AB}$ requires the same level of care as the previous
computation of $\gamma_{AB}$: the unperturbed expression $-r
\Omega_{AB}$ must be evaluated at $r = 2M(1+B)$ and $\theta^A =
\alpha^A + \Xi^A$ and combined with the terms that arise from the
metric perturbation.   

\subsection{Gauge transformations}
\label{subsec:gauge} 

We next work out how the various horizon quantities introduced
previously are affected by a gauge transformation of the form 
\begin{equation} 
x^a \to x^a + f^a, \qquad f^a = \eta^a(v,r) Y(\theta^A) 
\end{equation} 
and 
\begin{equation} 
\fl
\theta^A \to \theta^A + f^A, \qquad 
f^A = \Omega^{AB} \bigl[ \eta^{\rm even}(v,r) Y_B(\theta^A) 
+ \eta^{\rm odd}(v,r) X_B(\theta^A) \bigr]. 
\end{equation} 
We also have that $f_A = r^2 \Omega_{AB} f^B = (r^2 \eta^{\rm even})
Y_A + (r^2 \eta^{\rm odd}) X_A$. The gauge transformation affects the
coordinate description of the horizon. Recalling
Eq.~(\ref{hor_param}), we find that $b$, $\xi^{\rm even}$, and
$\xi^{\rm odd}$ change according to  
\begin{eqnarray} 
\Delta b &= (2M)^{-1} \eta_v(v,2M), \\
\Delta \xi^{\rm even} &= \eta^{\rm even}(v,2M), \\ 
\Delta \xi^{\rm odd} &= \eta^{\rm odd}(v,2M). 
\end{eqnarray} 
A complete listing of the corresponding changes in the metric
perturbation can be found in Sec.~\ref{subsec:metric}. 

With these rules it is easy to show that the quantities associated
with the horizon's intrinsic geometry change according to 
\begin{equation} 
\Delta \gamma^{\rm trace}= \Delta \gamma^{\rm even} 
= \Delta \gamma^{\rm odd} = 0. 
\end{equation} 
These results imply that $\gamma_{AB}$, $\Theta$, and $\sigma_{AB}$
are all gauge invariant, and we conclude that the horizon's intrinsic
geometry is gauge invariant. This is not a surprising conclusion. The
intrinsic metric is a collection of scalar fields with regards to
transformations of the spacetime coordinates $x^\alpha$, and it is
expressed entirely in terms of the intrinsic coordinates
$(\lambda,\alpha^A)$. As such it is as a matter of principle immune to
a gauge transformation. The fact that $\lambda$ is identified with the
spacetime coordinate $v$ adds a small complication to this argument,
because $\gamma_{AB}$ could in principle be sensitive to a change in
$v$. The identification associates a gauge transformation on $v$ to a
reparameterization of the generators, as was described in
Sec.~\ref{subsec:reparam}.  But $\gamma^0_{AB} = (2M)^2 \Omega_{AB}$,
the induced metric on the unperturbed horizon, is independent of $v$,
and the results displayed in
Eqs.~(\ref{repara_inf_a})--(\ref{repara_inf_e}) reveal that an 
infinitesimal reparameterization has no effect on $\delta
\gamma_{AB}$, the metric perturbation. The conclusion, therefore,
remains valid regardless of the identification $\lambda \equiv v$. As
an additional remark, we recall that the invariance of $\gamma_{AB}$
under general (large) reparameterizations was established in
Eqs.~(\ref{repara_a})--(\ref{repara_e}).   
  
On the other hand, the quantities associated with the horizon's
extrinsic geometry change according to 
\begin{eqnarray} 
\Delta k &= -4M \partial_v \bigl( \partial_v \eta_r 
+ \kappa_0 \eta_r \bigr), 
\label{extgeom_gauge_even} \\ 
\Delta \omega^{\rm even} &= -\bigl( \partial_v \eta_r 
+ \kappa_0 \eta_r \bigr), \\ 
\Delta {\cal K}^{\rm trace} &= -2M \partial_v \eta_r 
+ \frac{1}{2} \ell(\ell+1) \eta_r, \\ 
\Delta {\cal K}^{\rm even} &= -\eta_r, 
\end{eqnarray} 
and 
\begin{equation} 
\Delta \omega^{\rm odd} = 0 = \Delta {\cal K}^{\rm odd}. 
\label{extgeom_gauge_odd} 
\end{equation} 
In the even-parity sector, the changes in the extrinsic geometry are  
all associated with $\eta_r(v,2M) = \eta^v(v,2M)$, which describes a
change in $v$; there are no changes in the odd-parity sector. As
before we can observe that since $\kappa$, $\omega_A$, and
${\cal K}_{AB}$ are all spacetime scalars expressed entirely in terms
of the intrinsic coordinates, they should all be immune to a gauge
transformation. But as before we can identify a change in $v$ with a 
reparameterization of the generators, and infer from
Eqs.~(\ref{repara_inf_a})--(\ref{repara_inf_e}) the effect of the
reparameterization on the extrinsic geometry. With $\delta \lambda$
identified with $\eta_r(v,2M) Y(\alpha^A)$, we quickly find that
Eqs.~(\ref{repara_inf_a})--(\ref{repara_inf_e}) reproduce the
statements of
Eqs.~(\ref{extgeom_gauge_even})--(\ref{extgeom_gauge_odd}). 

It is easy to identify four linearly-independent quantities, formed
from $k$, $\omega^{\rm even}$, ${\cal K}^{\rm trace}$, and 
${\cal K}^{\rm even}$, that are invariant under infinitesimal
reparameterizations. We choose  
\begin{eqnarray} 
\psi_1 &:= \dot{\omega}^{\rm even} - \kappa_0 k, 
\label{gauge_inv_a} \\ 
\psi_2 &:= \dot{\cal K}^{\rm even} 
+ \kappa_0 {\cal K}^{\rm even} - \omega^{\rm even}, 
\label{gauge_inv_b} \\ 
\psi_3 &:= \dot{\cal K}^{\rm trace} 
+ \kappa_0 {\cal K}^{\rm trace} 
+ \frac{1}{2} \ell(\ell+1) \omega^{\rm even} - \frac{1}{2} k,
\label{gauge_inv_c} \\ 
\psi_4 &:= {\cal K}^{\rm trace} 
+ \frac{1}{2} \bigl[ \ell(\ell+1) + 1 \bigr] {\cal K}^{\rm even} 
- 2M \omega^{\rm even}. 
\label{gauge_inv_d} 
\end{eqnarray} 
The first three combinations can be shown to be pieces of the
spacetime Riemann tensor evaluated on the deformed horizon. Indeed,
inserting the results obtained in Sec.~\ref{subsec:extrinsic} within  
Eq.~(\ref{GCa}) yields 
\begin{equation} 
R_{\mu\nu\lambda\alpha} k^\mu N^\nu k^\lambda e^\alpha_A 
= \bigl( \dot{\omega}^{\rm even} - \kappa_0 k \bigr) Y_A 
+ \dot{\omega}^{\rm odd}\, X_A. 
\end{equation} 
It is easy to show that the left-hand side is invariant under
infinitesimal reparameterizations, and this guarantees that $\psi_1$
and $\omega^{\rm odd}$ also must be invariant. Similarly, we find from
Eq.~(\ref{GCRc}) that 
\begin{eqnarray} 
\fl
\frac{1}{2} R_{\alpha\beta} e^\alpha_A e^\beta_A &=
\biggl[ \dot{\cal K}^{\rm trace} 
+ \kappa_0 {\cal K}^{\rm trace} 
+ \frac{1}{2} \ell(\ell+1) \omega^{\rm even} - \frac{1}{2} k
\nonumber \\ \fl & \quad \mbox{}
+ \frac{1}{4} \ell(\ell+1) \gamma^{\rm trace} 
+ \frac{1}{8} (\ell-1)\ell(\ell+1)(\ell+2) \gamma^{\rm even} \biggr]
\Omega_{AB} Y 
\nonumber \\ \fl & \quad \mbox{}
+ \Bigl( \dot{\cal K}^{\rm even} 
+ \kappa_0 {\cal K}^{\rm even} - \omega^{\rm even} 
+ M \dot{\gamma}^{\rm even} + \frac{1}{2} \gamma^{\rm even} \Bigr)
Y_{AB} 
\nonumber \\ \fl & \quad \mbox{}
+ \Bigl( \dot{\cal K}^{\rm odd} 
+ \kappa_0 {\cal K}^{\rm odd} - \omega^{\rm odd} 
+ M \dot{\gamma}^{\rm odd} + \frac{1}{2} \gamma^{\rm odd} \Bigr)
X_{AB};
\end{eqnarray} 
invariance of $R_{\alpha\beta} e^\alpha_A e^\beta_B$ and $\gamma_{AB}$ 
under infinitesimal parameterizations guarantees that $\psi_2$ and
$\psi_3$ also must be invariant. The fourth quantity, $\psi_4$, does
not appear to be related in a similar way to a piece of the spacetime
Riemann tensor. 

\section{Tidal deformations} 
\label{sec:tidal} 

The formalism developed in the preceding section is very general, and
it can accommodate black-hole deformations created by matter flowing
across the event horizon, or by matter situated outside the black
hole's immediate neighbourhood. The formalism is also general relative
to the choice of gauge, because the relations between the horizon
quantities (such as $\gamma^{\rm trace}$, $\gamma^{\rm even}$, 
$\gamma^{\rm odd}$, $k$, $\omega^{\rm even}$, $\omega^{\rm odd}$, 
${\cal K}^{\rm trace}$, ${\cal K}^{\rm even}$, and 
${\cal K}^{\rm odd}$) and the metric perturbation are valid in any
gauge. How the horizon quantities change under gauge transformations 
(or better stated, reparameterizations of the horizon's null
generators) was described in Sec.~\ref{subsec:gauge}.  

In this section we specialize the situation to a tidal deformation of
a black hole created by a remote distribution of matter. We
incorporate the vacuum field equations into our analysis to relate the
horizon quantities to the well-known master functions 
$\Psi_{\rm even}$ and $\Psi_{\rm odd}$ of black-hole perturbation
theory (defined below). We next introduce a geometric notion of tidal
displacement on the event horizon, and describe how the tidal bulge is
related to the applied tidal field.  

\subsection{Master functions} 
\label{subsec:master} 

Gauge-invariant definitions of the master functions were provided in
Ref.~\cite{martel-poisson:05}. In the even-parity sector, 
$\Psi_{\rm even}$ is the Zerilli-Moncrief function 
\cite{zerilli:70, moncrief:74} defined by 
\begin{equation}
\Psi_{\rm even} := \frac{2r}{\lambda} \biggl[ \tilde{K}  
+ \frac{2}{\mu + 6M/r} \Bigl( r^a r^b \tilde{h}_{ab}  
- r r^a \nabla_a \tilde{K} \Bigr) \biggr],  
\end{equation}
where $\lambda := \ell(\ell+1) = \mu+2$, $\mu := (\ell-1)(\ell+2) 
= \lambda-2$, and where $\tilde{h}_{ab} := h_{ab} 
- \nabla_a \varepsilon_b - \nabla_b \varepsilon_a$, 
$\tilde{K} := K + \frac{1}{2} \lambda G - 2 r^a \varepsilon_a/r$, 
with $\varepsilon_a := j_a - \frac{1}{2} r^2 \nabla_a G$, are
gauge-invariant combinations of metric perturbations. We use the
notation $r_a := \partial r/\partial x^a$, $\nabla_a$ is the
covariant-derivative operator compatible with the two-dimensional
metric $g^0_{ab} dx^a dx^b = -f\, dv^2 + 2\, dvdr$, and as usual the
spherical-harmonic labels $\ell m$ are omitted. The
Zerilli-Moncrief function is known to satisfy the Zerilli
equation \cite{zerilli:70}, which is a two-dimensional wave equation
with an effective potential and a source term constructed from the
energy-momentum tensor of the matter distribution.  

In the odd-parity sector, $\Psi_{\rm odd}$ is the
Cunningham-Price-Moncrief function \cite{cunningham-etal:78,
  cunningham-etal:79} defined by  
\begin{equation}
\Psi_{\rm odd} := \frac{2r}{\mu} \varepsilon^{ab} 
\Bigl( \nabla_a \tilde{h}^{lm}_b - \frac{2}{r} r_a \tilde{h}^{lm}_b
\Bigr),  
\end{equation}
where $\varepsilon_{ab}$ is the Levi-Civita tensor on the
two-dimensional manifold with metric $g^0_{ab}$, and    
$\tilde{h}_a := h_a - \frac{1}{2} \nabla_a h_2 + r_a h_2/r$ is a
gauge-invariant combination of metric perturbations. The master
function is known to satisfy the Regge-Wheeler equation
\cite{regge-wheeler:57}, another two-dimensional wave equation with an
effective potential and a source term. The Regge-Wheeler equation is
also satisfied by another choice of master function, the original
Regge-Wheeler function \cite{regge-wheeler:57}; in vacuum this is
equal to half the time derivative of the Cunningham-Price-Moncrief
function.     

\subsection{Killing gauge} 
\label{subsec:killing} 

To relate the horizon quantities to $\Psi_{\rm even}$ and 
$\Psi_{\rm odd}$ it is convenient to adopt a ``Killing gauge'' defined
by 
\begin{equation} 
p_{\alpha\beta} t^\beta = 0, 
\label{killing_gauge}
\end{equation} 
where $t^\alpha$ is the timelike Killing vector of the Schwarzschild
spacetime. In the coordinates $(v,r,\theta,\phi)$ we have that
$t^\alpha = (1,0,0,0)$, and the gauge conditions translate to 
\begin{equation} 
h_{vv} = h_{vr} = j_{v} = 0
\end{equation} 
in the even-parity sector, and 
\begin{equation}  
h_v = 0
\end{equation} 
in the odd-parity sector. These conditions apply in a neighbourhood of 
the event horizon.  

An immediate virtue of the Killing gauge is that it preserves the
coordinate description of the event horizon, which continues, even in
the perturbed spacetime, to be described by $r=2M$ and $\theta^A =
\alpha^A$. In the terminology of Poisson and Vlasov
\cite{poisson-vlasov:10},  {\it the Killing gauge is a
  horizon-locking gauge}. This can be seen at once from
Eqs.~(\ref{teleo_solution1})--(\ref{teleo_sol_b}), which imply that 
\begin{equation} 
b(v) = \xi^{\rm even} = \xi^{\rm odd} = 0
\end{equation} 
whenever $h_{vv} = j_v = h_v = 0$ at $r=2M$.  We remark that while the
light-cone gauge adopted by Poisson and Vlasov also has the property
of being a horizon-locking gauge, the Killing gauge adopted here is
quite distinct from the light-cone gauge.   

\subsection{Near-horizon analysis} 

To calculate the horizon quantities we must integrate the perturbation 
equations in a neighbourhood of the event horizon; these are listed in
Secs.~IV B and V B of Ref.~\cite{martel-poisson:05}. In the even-parity
sector this can be accomplished by inserting the expansions 
$h_{rr} = h_0(v) + h_1(v)(r-2M) + h_2(v)(r-2M)^2 + \cdots$,
$j_{r} = j_0(v) + j_1(v)(r-2M) + j_2(v)(r-2M)^2 + \cdots$,
$K = K_0(v) + K_1(v)(r-2M) + K_2(v)(r-2M)^2 + \cdots$, and  
$G = G_0(v) + G_1(v)(r-2M) + G_2(v)(r-2M)^2 + \cdots$ within the
perturbation equations, and solving order-by-order in $r-2M$. Such an
analysis reveals that $j_0(v)$ is unconstrained by the field
equations, that $K_0(v) = 0$ and $G_0(v) = 2\mu^{-1} d j_0/dv$, and
that $K_1(v)$ and $G_1(v)$ must satisfy the differential equations 
\begin{equation} 
\frac{d K_1}{dv} + \kappa_0 K_1 = -4\lambda \kappa_0^3 j_0
\end{equation} 
and 
\begin{equation} 
\frac{d G_1}{dv} + \kappa_0 G_1 = 
-\frac{4 \kappa_0}{\mu} \biggl( \frac{d^2 j_0}{dv^2} 
- \mu \kappa_0 \frac{d j_0}{dv} - 2 \mu \kappa_0^2 j_0 \biggr),  
\end{equation} 
where $\kappa_0 := (4M)^{-1}$ is the surface gravity of the
unperturbed horizon. 

The fact that the differential operator acting on $K_1$ and $G_1$ is
$d/dv + \kappa_0$, instead of $d/dv - \kappa_0$ as in
Eq.~(\ref{horizon_equation}), implies that one should not look for
teleological solutions to these equations: the presence of
$e^{\kappa_0 v'}$ instead of $e^{-\kappa_0 v'}$ within the integrals 
would prevent them from converging if the integrations were
unbounded. We work instead with the most general solutions   
\begin{equation} 
K_1(v) = K_1(v_0) e^{-\kappa_0(v-v_0)} 
- 4\lambda \kappa_0^3 \int_{v_0}^v 
e^{-\kappa_0 (v-v')} j_0(v')\, dv' 
\end{equation} 
and 
\begin{eqnarray}
\fl 
G_1(v) &= \Biggl\{ G_1(v_0) + \frac{4\kappa_0}{\mu} \biggl[ 
\frac{dj_0}{dv}(v_0) - (\lambda-1) \kappa_0 j_0(v_0) \biggr] 
\Biggr\} e^{-\kappa_0(v-v_0)} 
\nonumber \\ \fl & \quad \mbox{} 
-\frac{4 \kappa_0}{\mu} \biggl[ 
\frac{d j_0}{dv}(v) - (\lambda-1) \kappa_0 j_0(v) 
- (\lambda-3) \kappa_0^2 \int_{v_0}^v 
e^{-\kappa_0 (v-v')} j_0(v')\, dv' \biggr], 
\end{eqnarray}   
in which the initial values $K_1(v_0)$ and $G_1(v_0)$ are not
determined by the requirements that $K_1(v\to \infty) \to 0$ and  
$G_1(v\to \infty) \to 0$. (We explore these issues further in
\ref{app:late-time} below.)  We do not need expressions for $h_0(v)$,
$h_1(v)$, $j_1(v)$ (nor other coefficients in the expansions), and the 
master function can be shown to be given by 
$\Psi_{\rm even} = (\mu\kappa_0)^{-1} d j_0/dv$ at $r=2M$.    

In the odd-parity sector we substitute the expansions 
$h_{r} = h_{r0}(v) + h_{r1}(v)(r-2M) + h_{r2}(v)(r-2M)^2 + \cdots$ and  
$h_2 = h_{20}(v) + h_{21}(v)(r-2M) + h_{22}(v)(r-2M)^2 + \cdots$ into
the perturbation equations, and solve order-by-order in $r-2M$. Such
an analysis reveals that $h_{r0}(v)$ is unconstrained by the field
equations, that $h_{20}(v) = (2\mu\kappa_0^2)^{-1} d h_{r0}/dv$, 
and that $h_{21}(v)$ satisfies the differential equation 
\begin{equation} 
\frac{d h_{21}}{dv} + \kappa_0 h_{21} = 
\frac{1}{\mu \kappa_0} \biggl( \frac{d^2 h_{r0}}{dv^2} 
+ \lambda \kappa_0 \frac{d h_{r0}}{dv} + 2 \mu \kappa_0^2 h_{r0}
\biggr).   
\end{equation} 
The general solution is 
\begin{eqnarray} 
\fl
h_{21}(v) &= \Biggl\{ h_{21}(v_0) - \frac{1}{\mu\kappa_0} \biggl[  
\frac{d h_{r0}}{dv}(v_0) + (\lambda-1) \kappa_0 h_{r0}(v_0) \biggr] 
\Biggr\} e^{-\kappa_0(v-v_0)} 
\nonumber \\ \fl & \hspace*{-20pt} \mbox{} 
+ \frac{1}{\mu \kappa_0} \biggl[ 
\frac{d h_{r0}}{dv}(v) + (\lambda-1) \kappa_0 h_{r0}(v)  
+ (\lambda-3) \kappa_0^2 \int_{v_0}^v 
e^{-\kappa_0 (v-v')} h_{r0}(v')\, dv' \biggr], 
\end{eqnarray}   
and the master function can be shown to be given by 
$\Psi_{\rm odd} = -(\mu\kappa_0)^{-1} d h_{r0}/dv$ at 
$r=2M$.   

\subsection{Horizon quantities} 
\label{subsec:hor_quantities} 

With the results obtained in the preceding subsection, we find that
the horizon quantities defined in Secs.~\ref{subsec:intrinsic} and
\ref{subsec:extrinsic} are given by 
\begin{eqnarray} 
\fl
\gamma^{\rm trace} &= 0, 
\label{horquant_a}\\ 
\fl
\gamma^{\rm even} &= 2\kappa_0 \Psi_{\rm even}(v,2M), \\
\fl
\gamma^{\rm odd} &= -2\kappa_0 \Psi_{\rm odd}(v,2M), \\
\fl
k &= 0, \\ 
\fl
\omega^{\rm even} &= \frac{1}{2} \mu \kappa_0 
  \Psi_{\rm even}(v,2M), \\ 
\fl
\omega^{\rm odd} &= -\frac{1}{2} \mu \kappa_0 
  \Psi_{\rm odd}(v,2M), \\  
\fl
{\cal K}^{\rm trace} &= {\cal K}^{\rm trace}(v_0) e^{-\kappa_0(v-v_0)} 
- \frac{1}{2} \lambda\mu \kappa_0 
  \int_{v_0}^v e^{-\kappa_0 (v-v')} \Psi_{\rm even}(v',2M)\, dv', \\
\fl
{\cal K}^{\rm even} &= \biggl[ {\cal K}^{\rm even}(v_0) 
+ \frac{1}{2} \Psi_{\rm even}(v_0,2M) \biggr] e^{-\kappa_0(v-v_0)} 
- \frac{1}{2} \Psi_{\rm even}(v,2M) 
\nonumber \\ \fl & \quad \mbox{} 
+ \frac{1}{2} (\lambda-3) \kappa_0 \int_{v_0}^v e^{-\kappa_0 (v-v')} 
\Psi_{\rm even}(v',2M)\, dv', \\   
\fl
{\cal K}^{\rm odd} &= \biggl[ {\cal K}^{\rm odd}(v_0) 
- \frac{1}{2} \Psi_{\rm odd}(v_0,2M) \biggr] e^{-\kappa_0(v-v_0)} 
+ \frac{1}{2} \Psi_{\rm odd}(v,2M) 
\nonumber \\ \fl & \quad \mbox{} 
- \frac{1}{2}(\lambda-3) \kappa_0 \int_{v_0}^v e^{-\kappa_0 (v-v')}  
\Psi_{\rm odd}(v',2M)\, dv',  
\label{horquant_b}
\end{eqnarray} 
where ${\cal K}^{\rm trace}(v_0)$, ${\cal K}^{\rm even}(v_0)$, and 
${\cal K}^{\rm odd}(v_0)$ can be expressed in terms of $K_1(v_0)$,
$G_1(v_0)$, and $j_0(v_0)$. We recall that 
$\lambda := \ell(\ell+1) = \mu+2$ and 
$\mu := (\ell-1)(\ell+2) = \lambda-2$.  

The results display a pleasing symmetry
(up to signs, which are inherited from the definitions of the master 
functions) between the even-parity and odd-parity sectors. In the case
of the intrinsic-geometry quantities $\gamma^{\rm even}$ and 
$\gamma^{\rm odd}$, the symmetry is gauge
invariant; in the case of the extrinsic-geometry quantities  
$\omega^{\rm even}$ and $\omega^{\rm odd}$, ${\cal K}^{\rm even}$ and   
${\cal K}^{\rm odd}$, the symmetry is a property of the Killing gauge 
adopted here (it is not, in particular, a property of the light-cone
gauge \cite{poisson-vlasov:10}). Another remarkable property of the
Killing gauge is the fact that $k = 0$, so that the surface gravity of
the perturbed black hole is $\kappa = \kappa_0 = (4M)^{-1}$. 
  
The expressions for ${\cal K}^{\rm trace}$ and ${\cal K}^{\rm even}$
given previously were simplified relative to the more primitive
expressions obtained in terms of $j_0$. These, however,  involved the
combination  
\begin{equation} 
\fl
j_0 - \kappa_0 \int_{v_0}^v e^{-\kappa(v-v')} j_0(v')\, dv' 
= j_0(v_0) e^{-\kappa_0(v-v_0)} 
+ \int_{v_0}^v e^{-\kappa(v-v')} \frac{d j_0}{d v'}\, dv', 
\end{equation} 
which could readily be expressed in terms of $\Psi_{\rm even}  
= (\mu\kappa_0)^{-1} d j_0/dv$. A very similar simplification was
achieved in the case of ${\cal K}^{\rm odd}$. 

The gauge-invariant quantities defined by
Eqs.~(\ref{gauge_inv_a})--(\ref{gauge_inv_d}) are easily shown to be
given by 
\begin{eqnarray} 
\fl
\psi_1 &= \frac{1}{2} \mu \kappa_0 
 \partial_v \Psi_{\rm even}(v,2M), \\ 
\fl
\psi_2 &= -\frac{1}{2} \partial_v \Psi_{\rm even}(v,2M)
- \kappa_0 \Psi_{\rm even}(v,2M),\\ 
\fl
\psi_3 &= -\frac{1}{4} \lambda\mu \kappa_0 \Psi_{\rm even}(v,2M), \\ 
\fl
\psi_4 &= \biggl[ {\cal K}^{\rm trace}(v_0) 
+ \frac{1}{2}(\lambda+1) {\cal K}^{\rm even}(v_0) 
+ \frac{1}{4}(\lambda+1) \Psi_{\rm even}(v_0,2M) \biggr]
e^{-\kappa_0(v-v_0)} 
\nonumber \\ \fl & \quad \mbox{} 
-\frac{1}{4}(2\lambda-1) \Psi_{\rm even}(v,2M) 
- \frac{1}{4} (\lambda\mu + 3) \kappa_0 
\int_{v_0}^v e^{-\kappa_0 (v-v')} \Psi_{\rm even}(v',2M)\, dv'. 
\end{eqnarray} 

\subsection{Intrinsic geometry and tidal displacement}   
\label{subsec:bulge} 

The results obtained in the preceding subsection imply that the
induced metric on the event horizon simplifies to 
\begin{eqnarray}
\gamma_{AB} &= (2M)^2 \bigl( \Omega_{AB} 
+ \delta \gamma_{AB} \bigr), 
\label{mt_a} \\ 
\delta \gamma_{AB} &= 2\kappa_0 \bigl[  
\Psi_{\rm even}(v,2M) Y_{AB} 
- \Psi_{\rm odd}(v,2M) X_{AB} \bigr] 
\label{mt_b}
\end{eqnarray}  
in the case of a tidally deformed black hole. From
Eqs.~(\ref{expansion}) and (\ref{shear}) we also get 
\begin{equation} 
\Theta = 0
\label{expansion_tidal} 
\end{equation} 
and 
\begin{equation} 
\sigma_{AB} = M \bigl[ \partial_v \Psi_{\rm even}(v,2M) Y_{AB}  
- \partial_v \Psi_{\rm odd}(v,2M) X_{AB} \bigr].  
\label{shear_tidal} 
\end{equation} 
The fact that the expansion vanishes to leading order in perturbation
theory can be derived directly from Raychaudhuri's equation: The
reduction of Eq.~(\ref{ray1}) to vacuum and to first-order
perturbation theory is $\partial_v \Theta = \kappa_0 \Theta$, and this
implies (with an appropriate choice of final condition) that $\Theta$
must vanish.  

The reduction of Eq.~(\ref{ray2}) gives an expression for the Weyl
tensor evaluated on the event horizon: 
\begin{equation} 
C_{AB} = (\kappa_0 - \partial_v) \sigma_{AB}, 
\label{weyl_tidal} 
\end{equation} 
where $C_{AB} := C_{\mu\alpha\nu\beta} k^\mu e^\alpha_{(A)} k^\nu
e^\beta_{(B)}$. This equation can be integrated to relate the shear
tensor to the Weyl tensor; the appropriate teleological solution is 
\begin{equation} 
\sigma_{AB}(v,\alpha^A) = \int_v^\infty e^{-\kappa_0(v'-v)} 
C_{AB}(v',\alpha^A)\, dv'. 
\label{shear_weyl} 
\end{equation} 
This equation implies that the shear tensor anticipates the 
behaviour of the Weyl tensor by a time interval of order
$\kappa_0^{-1} = 4M$. If the Weyl tensor is identified with the tidal 
field acting on the black hole, and if the shear tensor is adopted as
a measure of tidal deformation, then we have the statement that
{\it the tide leads the applied field by a time interval 
of order $4M$}. This observation was already made by Fang and Lovelace
\cite{fang-lovelace:05} in a more restricted context (and by Hartle
\cite{hartle:74} in the case of a rotating black hole), and we find here
that it holds in all generality as a consequence of the teleological
nature of the event horizon. We remark that in the case of a Newtonian
body made up of a viscous fluid, the tide would be lagging instead of
leading (when the body is nonrotating), and that the time interval
would be proportional to $R \nu/M$, with $R$ denoting the body's
averaged radius, $\nu$ its kinematic viscosity, and $M$ its mass.    

Another meaningful measure of tidal deformation comes from the
Ricci curvature scalar associated with the metric of
Eqs.~(\ref{mt_a}) and (\ref{mt_b}). This is given by
Eq.~(\ref{ricci}) with $\gamma^{\rm trace} = 0$ and 
$\gamma^{\rm even} = 2\kappa_0 \Psi_{\rm even}(v,2M)$:
\begin{equation} 
{\cal R} = \frac{1}{2M^2} \biggl[ 1 + \frac{1}{2} (\ell-1)\ell(\ell+1) 
(\ell+2) \kappa_0 \Psi_{\rm even}(v,2M) Y(\alpha^A) \biggr]. 
\label{ricci_tidal} 
\end{equation} 
It is helpful to convert this into a dimensionless {\it tidal
displacement field} $\rho(v,\alpha^A)$ by identifying ${\cal R}$ with 
the curvature of a two-dimensional surface embedded in a flat,
three-dimensional space. We describe this surface in spherical
coordinates $(r,\alpha^A)$ by the parametric equation 
$r = 2M[1 + \rho(v,\alpha^a)]$ with $\rho = \epsilon(v) Y(\alpha^A)$,
and demand that its curvature be equal to ${\cal R}$. We thus obtain  
$2M^2 {\cal R} = [ 1 + (\ell-1)(\ell+2) \epsilon Y(\alpha^A)]$,    
and the identification 
\begin{equation} 
\rho(v,\alpha^A) = 
\frac{1}{2} \ell(\ell+1) \kappa_0 \Psi_{\rm even}(v,2M) Y(\alpha^A)
\label{tidal_displacement} 
\end{equation} 
follows immediately. Once more summation over the omitted
spherical-harmonic labels $\ell m$ is understood.

The absence of trace terms in Eqs.~(\ref{mt_b}),
(\ref{shear_tidal}), and (\ref{weyl_tidal}) implies that each tensor   
$\delta \gamma_{AB}$, $\sigma_{AB}$, and $C_{AB}$ possesses only two      
independent components. Introducing the basis vectors 
\begin{equation} 
\upalpha^A = [1,0], \qquad 
\upbeta^A = [1,1/\sin\alpha] 
\label{angular_basis} 
\end{equation} 
in the horizon coordinates $\alpha^A = (\alpha,\beta)$, we take the
independent components of $\delta\gamma_{AB}$ to be 
\begin{eqnarray}
\gamma_+ &:= \frac{1}{2} \bigl( \upalpha^A \upalpha^B 
- \upbeta^A \upbeta^B \bigr) \delta \gamma_{AB}, 
\label{pol_a}
\\ 
\gamma_\times &:= \frac{1}{2} \bigl( \upalpha^A \upbeta^B 
+ \upbeta^A \upalpha^B \bigr) \delta \gamma_{AB}. 
\label{pol_b}
\end{eqnarray} 
The independent components $\sigma_{+,\times}$ and $C_{+,\times}$ of
the shear and Weyl tensors are defined in a similar manner. These
quantities are closely analogous to the gravitational-wave
polarizations $h_{+,\times}$ that can be defined in the wave zone of
an asymptotically-flat spacetime.  

\section{Applications} 
\label{sec:applications} 

\subsection{Slowly-varying quadrupolar tidal field}  
\label{subsec:quadrupole} 

As an application of the general formalism developed here we revisit
the situation examined by Poisson and Vlasov \cite{poisson-vlasov:10},
that of a black hole deformed by a slowly-varying tidal field. To
simplify the discussion we neglect the nonlinear terms included in
Ref.~\cite{poisson-vlasov:10}, and we specialize the tidal field to a
pure quadrupolar form.   

As described in Sec.~II of Ref.~\cite{poisson-vlasov:10}, the black
hole's tidal environment is described by the {\it tidal moments} 
$\E_{jk}(v)$ and $\B_{jk}(v)$. These quantities are
symmetric-tracefree (STF) Cartesian tensors that represent the
components of the spacetime Weyl tensor evaluated far away from the
black hole; latin indices $j$ and $k$ (and so on) run over the
values $1$, $2$, and $3$. The tidal moments give rise to the 
{\it tidal potentials}  
\begin{eqnarray} 
\EE{q} &:= \E_{pq} \Omega^p \Omega^q, \\ 
\EE{q}_j &:= P_j^{\ p} \E_{pk} \Omega^k, \\
\EE{q}_{jk} &:= 2 P_j^{\ p} P_k^{\ q} \E_{pq} + P_{jk} \EE{q} 
\end{eqnarray} 
and 
\begin{eqnarray} 
\BB{q}_j &:= \epsilon_{jpq} \Omega^p \B^q_{\ n} \Omega^n, \\ 
\BB{q}_{jk} &:= \epsilon_{jpq} \Omega^p B^q_{\ n} P^n_{\ k}
+ \epsilon_{kpq} \Omega^p B^q_{\ n} P^n_{\ j}, 
\end{eqnarray} 
where the label ``${\sf q}$'' stands for ``quadrupolar,'' $\Omega^j := 
[\sin\alpha\cos\beta, \sin\alpha\sin\beta, \cos\alpha]$ is a Cartesian 
unit vector constructed from the generator labels 
$\alpha^A = (\alpha,\beta)$, $P_{jk} := \delta_{jk} - \Omega_j 
\Omega_k$ is a projection operator to the subspace transverse to
$\Omega^j$, and $\epsilon_{jkn}$ is the Cartesian permutation
symbol. The vector potentials $\EE{q}_j$ and $\BB{q}_j$ are
transverse, in the sense that $\EE{q}_j \Omega^j = 0 
= \BB{q}_j \Omega^j$. In addition to being transverse, the tensor
potentials $\EE{q}_{jk}$ and $\BB{q}_{jk}$ are also tracefree, in the
sense that $\delta^{jk} \EE{q}_{jk} = 0 = \delta^{jk} \BB{q}_{jk}$. In
all manipulations involving Cartesian tensors, latin indices are
lowered and raised with the Euclidean metric $\delta_{jk}$.  

The vectorial and tensorial potentials can be converted to angular
components by means of the transformation matrix $\Omega^j_A
:= \partial \Omega^j/\partial\alpha^A$. We thus introduce 
\begin{equation} 
\EE{q}_A := \EE{q}_j \Omega^j_A, \qquad 
\EE{q}_{AB} := \EE{q}_{jk} \Omega^j_A \Omega^k_B 
\end{equation} 
and 
\begin{equation} 
\BB{q}_A := \BB{q}_j \Omega^j_A, \qquad 
\BB{q}_{AB} := \BB{q}_{jk} \Omega^j_A \Omega^k_B. 
\end{equation} 
As shown in Sec.~II of Ref.~\cite{poisson-vlasov:10}, these angular
potentials can be expressed as expansions in spherical harmonics of
degree $\ell = 2$. We have  
\begin{equation} 
\fl
\EE{q} =\sum_m \E_m Y^{2,m}, \qquad
\EE{q}_A = \frac{1}{2} \sum_m \E_m Y^{2,m}_A, \qquad 
\EE{q}_{AB} = \sum_m \E_m Y^{2,m}_{AB} 
\label{E_sphe} 
\end{equation} 
and 
\begin{equation} 
\BB{q}_A = \frac{1}{2} \sum_m \B_m X^{2,m}_A, \qquad 
\BB{q}_{AB} = \sum_m \B_m X^{2,m}_{AB}.  
\label{B_sphe} 
\end{equation} 
The sums are carried out from $m=-2$ to $m=2$, the coefficients $\E_m$ 
and $\B_m$ are related to $\E_{jk}$ and $\B_{jk}$ and depend on $v$
only; the spherical harmonics are functions of $\alpha^A$. These
expansions reveal that $\E_{jk}(v)$ gives rise to a perturbation of
even parity, while $\B_{jk}(v)$ gives rise to a perturbation of odd
parity.  

Solutions to the perturbation equations corresponding to a black hole
deformed by a quadrupolar tidal field were constructed by Poisson and
Vlasov \cite{poisson-vlasov:10}. The construction assumes that the
tidal moments vary slowly, in the sense that the timescale $\tau$
associated with these variations (denoted ${\cal R}$ in
Ref.~\cite{poisson-vlasov:10}) is very long compared with the
black-hole mass. The solutions were provided in the  
light-cone gauge, but it is easy from these results to obtain the
gauge-invariant master functions. The relations, in fact, are the same
as in the Killing gauge adopted in Sec.~\ref{subsec:killing}: we have
that $\Psi_{\rm even}(v,2M) = 2M G(v,2M)$ and   
$\Psi_{\rm odd}(v,2M) = - (2M)^{-1} h_2(v,2M)$, where $G$ and $h_2$
are obtained in the light-cone gauge. Importing the results of
Ref.~\cite{poisson-vlasov:10} --- summarized in their Eqs.~(6.10),
(6.18) and Table XIV --- we find that $G(v,2M) = -\frac{2}{3} M^2
\E_m$ and $h_2(v,2M) = -\frac{8}{3} M^4 \B_m$, so that 
\begin{eqnarray} 
\Psi_{\rm even}(v,2M) &= -\frac{4}{3} M^3 \E_m(v) 
\Bigl[ 1 + O(M^3/\tau^3) \Bigr], 
\label{master_quad_a} \\
\Psi_{\rm odd}(v,2M) &= \frac{4}{3} M^3 \B_m(v)
\Bigl[ 1 + O(M^3/\tau^3) \Bigr]. 
\label{master_quad_b} 
\end{eqnarray} 
Notice that these expressions involve $\E_m$ and $\B_m$ only, and not
their derivatives with respect to $v$, which would contribute
fractional corrections of order $M/\tau$ and $(M/\tau)^2$. As
explained in Sec.~VI A of Ref.~\cite{poisson-vlasov:10}, this property
results from the freedom to redefine the tidal moments according to 
$\E_{jk} \to \E_{jk} + p_1 M \dot{\E}_{jk} 
+ p_2 M^2 \ddot{\E}_{jk} + \cdots$ and $\B_{jk} \to \B_{jk} 
+ q_1 M \dot{\B}_{jk} + q_2 M^2 \ddot{\B}_{jk} + \cdots$, where $p_1$,
$p_2$, $q_1$, and $q_2$ are arbitrary numbers. 
 
It is a simple matter to insert Eqs.~(\ref{master_quad_a}) and
(\ref{master_quad_b}) within
Eqs.~(\ref{horquant_a})--(\ref{horquant_b}) and to calculate the
horizon quantities. Because the tidal moments $\E_{jk}$ and $\B_{jk}$
vary slowly, the integrations can be carried out as in
\ref{app:late-time}, by repeated integration by parts. After
discarding the transient terms that decay exponentially, we arrive at  
\begin{eqnarray} 
\gamma^{\rm trace} &= 0, \\ 
\gamma^{\rm even} &= -\frac{2}{3} M^2 \E_m, \\ 
\gamma^{\rm odd} &= -\frac{2}{3} M^2 \B_m, \\ 
k &= 0, \\ 
\omega^{\rm even} &= -\frac{2}{3} M^2 \E_m, \\ 
\omega^{\rm odd} &= -\frac{2}{3} M^2 \B_m, \\ 
{\cal K}^{\rm trace} &= 16 M^3 \bigl( \E_m - 4 M \dot{\E}_m 
+ 16 M^2 \ddot{\E}_m \bigr), \\ 
{\cal K}^{\rm even} &= -\frac{4}{3} M^3 \bigl( \E_m 
- 6 M \dot{\E}_m + 24 M^2 \ddot{\E}_m \bigr), \\ 
{\cal K}^{\rm odd} &= -\frac{4}{3} M^3 \bigl( \B_m 
- 6 M \dot{\B}_m + 24 M^2 \ddot{\B}_m \bigr). 
\end{eqnarray}
These expressions are valid up to correction terms of fractional order
$(M/\tau)^3$; they are given in the Killing gauge introduced in
Eq.~(\ref{killing_gauge}). 

These results, together with Eqs.~(\ref{E_sphe}) and (\ref{B_sphe}), 
imply that in the Killing gauge, 
\begin{eqnarray} 
\fl
\gamma_{AB} &= (2M)^2 \biggl[ \Omega_{AB} 
- \frac{2}{3} M^2 \bigl( \EE{q}_{AB} + \BB{q}_{AB} \bigr)\biggr], \\ 
\fl
\kappa &= \frac{1}{4M}, \\ 
\fl
\omega_A &= -\frac{4}{3} M^2 \bigl( \EE{q}_{AB} + \BB{q}_{AB} \bigr), \\ 
\fl
{\cal K}_{AB} &= -2M \Omega_{AB}  
+ 16M^3 \bigl( \EE{q} - 4M \dEE{q} + 16 M^2 \ddEE{q} \bigr)
\Omega_{AB} 
\nonumber \\ \fl & \quad \mbox{}
- \frac{4}{3} M^3 \biggl[ \bigl( \EE{q}_{AB} + \BB{q}_{AB} \bigr) 
- 6M \bigl( \dEE{q}_{AB} + \dBB{q}_{AB} \bigr) 
+ 24M^2 \bigl( \ddEE{q}_{AB} + \ddBB{q}_{AB} \bigr) \biggr]; 
\end{eqnarray} 
as before these expressions are accurate up to terms involving the
third derivative of the tidal moments. We showed in 
Sec.~\ref{subsec:gauge} that $\gamma_{AB}$ is gauge-invariant, while
$\kappa$, $\omega_A$, and ${\cal K}_{AB}$ are affected by a
reparameterization of the horizon's null generators. Gauge-invariant
combinations of these quantities were identified, and in particular we
have that  
\begin{equation} 
R_{\mu\nu\lambda\alpha} k^\mu N^\nu k^\lambda e^\alpha_A 
= -\frac{4}{3} M^2 \bigl( \dEE{q}_A + \dBB{q}_A \bigr) 
\end{equation} 
is invariant under infinitesimal reparameterizations; because it
originates from $\dot{\omega}^{\rm even}$ and 
$\dot{\omega}^{\rm even}$, this expression is accurate up to the
fourth derivative of the tidal moments. 

\subsection{Parabolic encounter} 
\label{subsec:parabolic} 

\begin{figure} 
\includegraphics[width=0.7\linewidth]{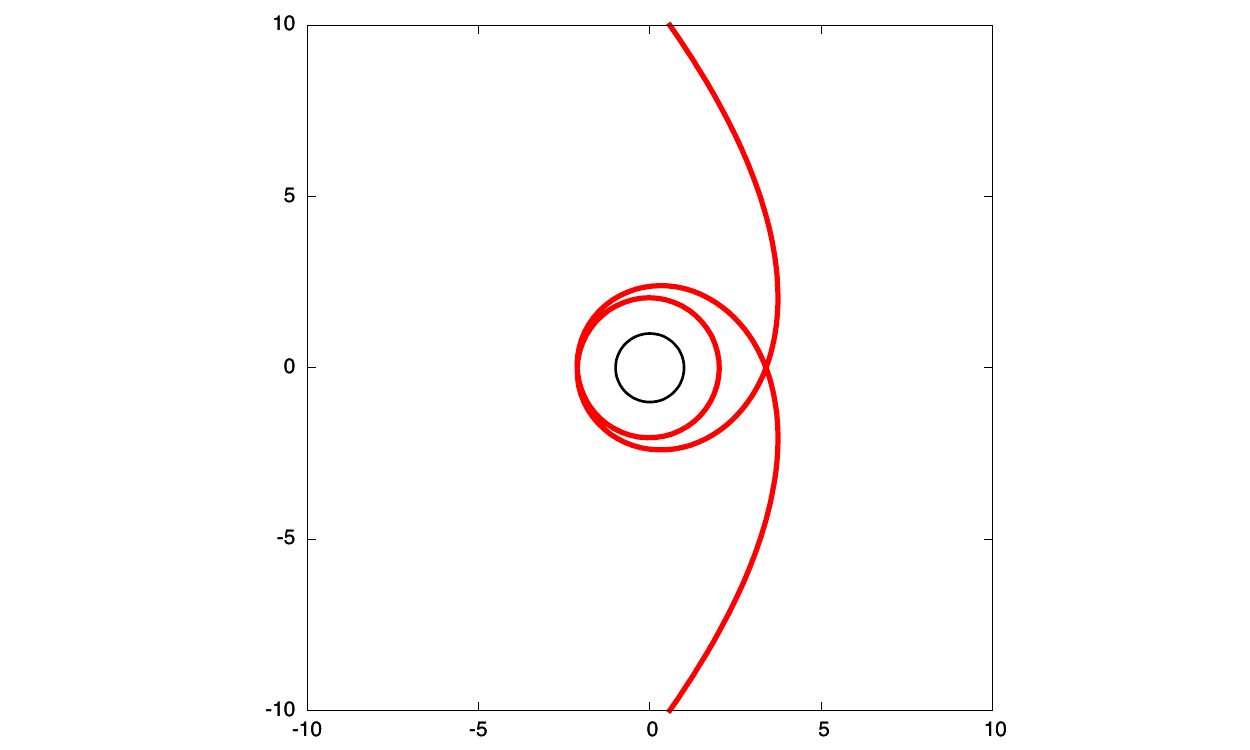}
\caption{Orbit of a parabolic encounter between a small body of mass
  $m$ and a black hole of mass $M$. The orbit's semi-latus rectum is
  $p = 8.1 M$ and its eccentricity is $e = 1$. The particle begins
  from rest at infinity, reaches a radial turning point at $r = 4.05
  M$, and returns to rest at infinity. The orbit is displayed in a
  $x$-$y$ plane constructed in the usual way from the Schwarzschild
  coordinates $r$ and $\phi$, so that $x = r\cos\phi$ and 
  $y = r\sin\phi$. The coordinates are rescaled by a factor of $2M$ to
  make them dimensionless; in these units the unperturbed horizon
  (shown in black) is described by a circle of unit radius. The
  orbital motion is calibrated so that $\phi = 0$ when 
  $r = 4.05M$.}  
\label{fig:orbit} 
\end{figure}

As a second application of the formalism we consider a parabolic
encounter between a particle of mass $m$ and a black hole of mass
$M$. We take $m$ to be much smaller than $M$, and we take the motion
of the particle to be a geodesic in the Schwarzschild spacetime. We
give the orbit a semi-latus rectum $p = 8.1 M$ and an eccentricity 
$e = 1$;  the parameterization is such that the radial turning points
are situated at $r_{\rm min} = p/(1+e) = 4.05M$ and 
$r_{\rm max} = p/(1-e) = \infty$. The orbit has a Killing energy 
$E = m$ and a Killing angular momentum 
$L \simeq 4.0003 mM$. The particle begins from rest at infinity, moves
inward, circles approximately twice around the black hole, moves
outward, and returns to rest at infinity; the shape of the orbit is
displayed in Fig.~\ref{fig:orbit}. Because the turning point is
so close to the black hole, the motion is highly relativistic when the
particle revolves around the black hole, and the tidal interaction is
highly dynamical.  

We calculate the gravitational perturbations created by the orbiting
particle by integrating the Zerilli and Regge-Wheeler equations for
the master functions $\Psi_{\rm even}$ and $\Psi_{\rm odd}$. This must
be accomplished numerically, and we rely on the time-domain,
finite-difference code written by Karl Martel; the details of the code
are described in Refs.~\cite{martel-poisson:02, martel:04}. 
Martel's original code had to be modified to account for a
different choice of odd-parity master function: While Martel's code
integrates the Regge-Wheeler equation for the original Regge-Wheeler
function (which is equal to $\frac{1}{2} \partial_t \Psi_{\rm even}$),
our modified version of the code calculates instead the
Cunningham-Price-Moncrief function $\Psi_{\rm even}$. The code returns
the master functions evaluated as functions of $v$ at a fixed radial
position $r = 2M(1+\epsilon)$ close to the event horizon; in our runs
we chose $\epsilon \simeq 10^{-5}$.  

\begin{figure} 
\includegraphics[width=0.7\linewidth]{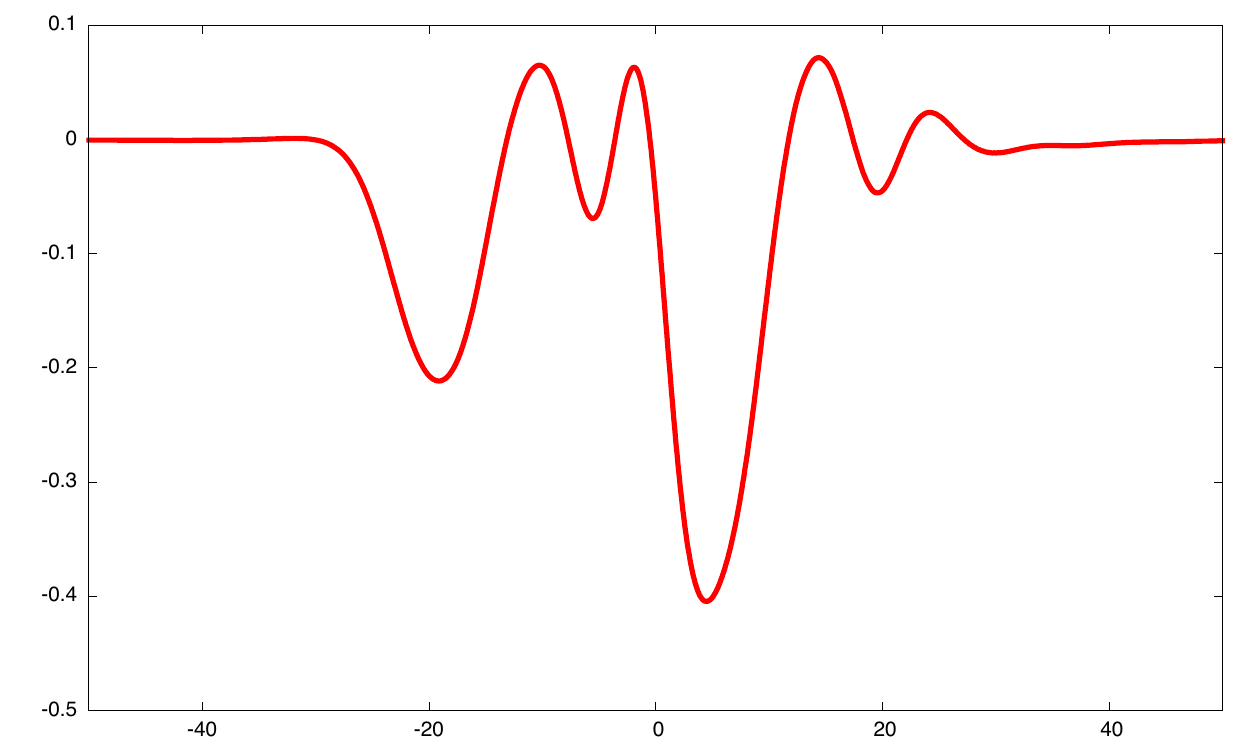}
\caption{Polarization $\gamma_+$ associated with the intrinsic
  geometry of a black-hole horizon perturbed by a parabolic
  encounter, calculated at azimuthal position $\beta = 0$ on the 
  horizon's equatorial plane $\alpha = \frac{\pi}{2}$, which coincides
  with the orbital plane. The polarization is displayed as a function
  of $v/(2M)$ and is rescaled by a factor of $m/(2M)$. All relevant
  multipoles up to $\ell = 4$ are included in the computation.}  
\label{fig:pol} 
\end{figure} 
 
In Fig.~\ref{fig:pol} we plot the
polarization $\gamma_+$ associated with the horizon's intrinsic 
geometry, as defined by Eq.~(\ref{pol_a}); this is shown as a function
of advanced-time $v$ at azimuthal position $\beta = 0$ on the orbital
plane $\alpha = \frac{\pi}{2}$; for this orientation we have that 
$\gamma_\times = 0$. The calculation involves a summation over all
multipoles up to (and including) $\ell = 4$; multipoles with 
$\ell \geq 5$ give contributions that are too small to be visible in
the plot. Most of the signal is produced when the particle revolves
around the black hole, and the plot reveals the rich harmonic
structure that a parabolic encounter imprints on the tidal deformation
of an event horizon.    

\begin{figure} 
\includegraphics[width=0.7\linewidth]{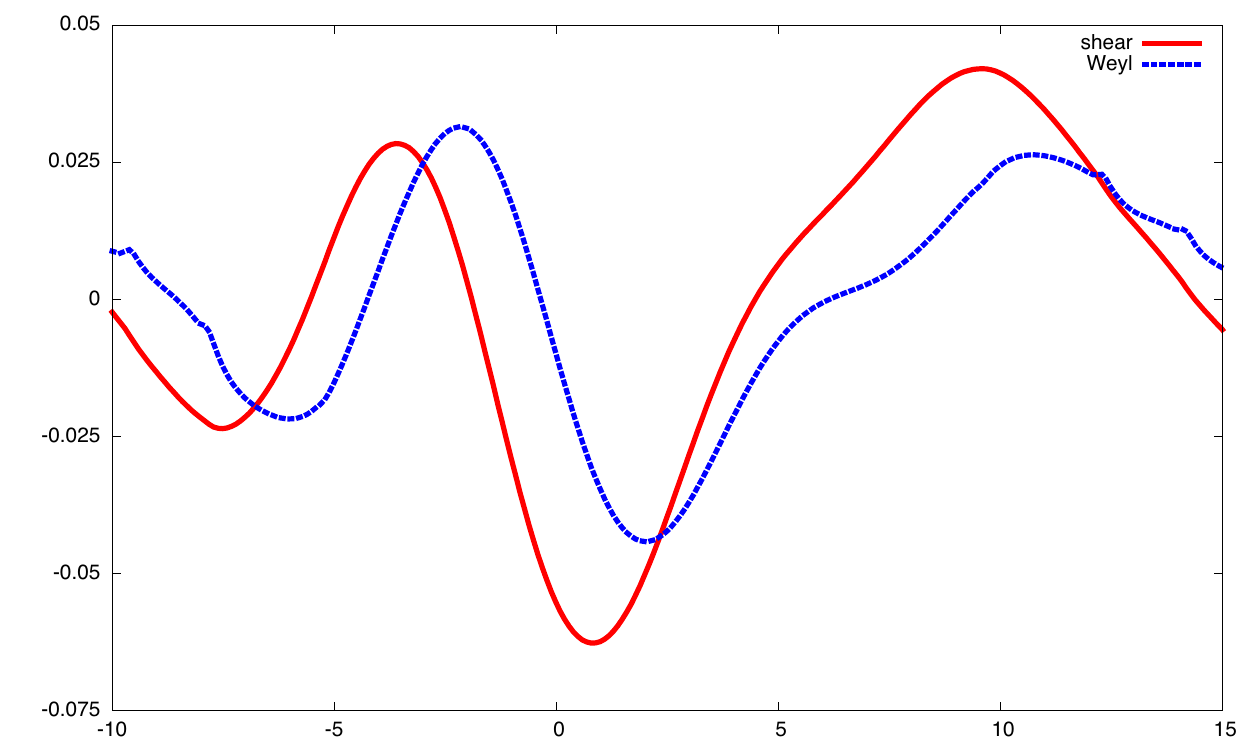}
\caption{Shown in solid red is the polarization $\sigma_+$ associated
  with the shear tensor of the horizon's intrinsic geometry, rescaled
  by a factor of $m$, as a function of $v/(2M)$. Shown in dashed blue
  is the polarization $C_+$ associated with the Weyl tensor, rescaled
  by a factor of $m/(2M)$, as a function of $v/(2M)$. Both quantities
  are calculated at position $\alpha = \frac{\pi}{2}$ and $\beta = 0$
  on the horizon. The plots show clearly that the horizon tide (as
  measured by the shear tensor) {\it leads} the tidal field (as
  measured by the Weyl tensor) by a time interval of order 
  $\kappa^{-1}_0 = 4M$. The Weyl tensor is noisy for early and late
  times because it is inaccurately computed by estimating the
  second derivative of $\gamma_+$ with respect to $v$ with
  finite-difference techniques.}  
\label{fig:shear} 
\end{figure} 

In Fig.~\ref{fig:shear} we plot the polarizations $\sigma_+ =
\frac{1}{2} \partial_v \gamma_+$ and $C_+ = (\kappa_0  
- \partial_v)\sigma_+$ of the shear and Weyl tensors, respectively;
these also are displayed as functions of $v$ at position 
$\alpha = \frac{\pi}{2}$ and $\beta = 0$ on the event horizon. The
figure reveals very clearly that the horizon tide (as measured by the
shear tensor) {\it leads} the tidal field (as measured by the Weyl
tensor) by a time interval of order $\kappa^{-1}_0 = 4M$; this
feature of the tidal dynamics of a nonrotating black hole was
discussed in Sec.~\ref{subsec:bulge}.   

\begin{figure} 
\includegraphics[width=0.7\linewidth]{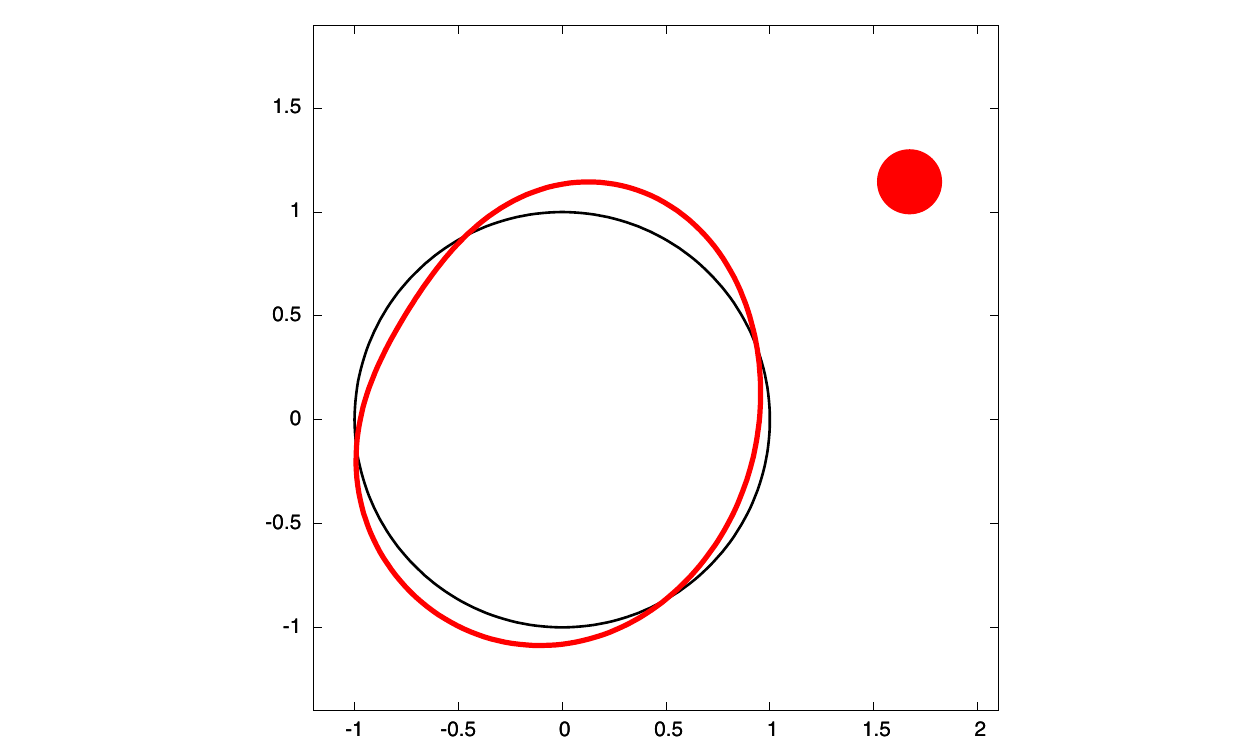}
\caption{Snapshot of the tidal bulge at $v/(2M) = 4.4874$ as
  described by the dimensional tidal displacement $\rho$ evaluated as
  a  function of $\beta$ on the black hole's equatorial plane 
  $\alpha = \frac{\pi}{2}$. The figure shows, in the same
  $x$-$y$ plane as in Fig.~1, the surface $r = 2M$ of the unperturbed
  horizon (in thin black) as well as the surface $r = 2M[1 + (M/m) \rho]$ (in
  thick red), which grossly exaggerates the horizon deformation by a
  factor of $M/m$ to make it visible. The figure also shows (red disk)
  the position of the orbiting body at this value of advanced time
  $v$; we have $r/(2M) \simeq 2.0273$ and $\phi \simeq 0.6005$,
  leading to the Cartesian positions $x/(2M)  \simeq 1.6726$ and 
  $y/(2M) \simeq 1.1455$. The tidal bulge and orbiting body are
  intersected by the same light cone $v = \mbox{constant}$, and here
  also we see the tidal bulge leading the source of the tidal field.}   
\label{fig:bulge} 
\end{figure}

Finally, in Fig.~\ref{fig:bulge} we display the shape of the tidal 
bulge at a selected value of $v$ in relation to the position of the
orbiting body. The tidal bulge is described geometrically in terms of
the tidal displacement field $\rho(v,\alpha^A)$ defined by 
Eq.~(\ref{tidal_displacement}), and the body's position is evaluated 
on the past light cone $v = \mbox{constant}$ so as to yield a
meaningful comparison. Here also we find that the horizon tide (as
measured by the displacement field) {\it leads} the source of the tide
(as measured by the orbital position on the light cone).   

\ack
This work was supported by the Natural Sciences and Engineering
Research Council of Canada.

\appendix
\section{Late-time behaviour of horizon quantities} 
\label{app:late-time} 

Some of the horizon quantities (such as $\gamma^{\rm even}$,
$\gamma^{\rm odd}$, $\omega^{\rm even}$, and 
$\omega^{\rm odd}$) can be expressed purely in terms of the current
value of the master functions, while others (such as 
${\cal K}^{\rm trace}$, ${\cal K}^{\rm even}$, and 
${\cal K}^{\rm odd}$) involve integrals of the master functions. We
wish to verify that all horizon quantities properly vanish at
$v=\infty$, assuming that $\Psi_{\rm even}(v,2M)$ and 
$\Psi_{\rm odd}(v,2M)$ decay at least as fast as an inverse power law
in $v$; this is the late-time behaviour expected of radiative tails
that linger on after the external processes that produce the
perturbation have shut down.  

The general structure of the integrals is 
\begin{equation} 
x(v) = x(v_0) e^{-\kappa_0(v-v_0)} - \int_{v_0}^v e^{-\kappa_0(v-v')} 
F(v')\, dv',
\end{equation} 
and for our purposes here we assume that the source function $F(v)$
varies over a timescale $\tau$ that is very long compared with 
$\kappa_0^{-1} = 4M$. In the case of an inverse-power falloff, for
example, we assume that $v_0$ is sufficiently large that 
$F(v') \propto (v')^{-p}$ within the integral, with $p > 0$. 
Then $\dot{F} \propto (v')^{-p-1}$ and the timescale
$\tau$ can be identified with $F/\dot{F} \propto v'$; this is indeed
much larger than $4M$ for the specified domain of integration. In these
circumstances we can evaluate the integral and express it as an
asymptotic series in powers of $(\kappa_0 \tau)^{-1} \ll 1$. If we let 
\begin{equation} 
I[F] :=  \int_{v_0}^v e^{-\kappa_0(v-v')} F(v')\, dv',
\end{equation} 
then the identity 
\begin{equation} 
I[F] = \frac{1}{\kappa_0} \biggl\{ F(v) - F(v_0) e^{-\kappa_0 (v-v_0)} 
- I[\dot{F}] \biggr\} 
\end{equation} 
follows immediately by integration by parts. Repeated applications
yield 
\begin{eqnarray} 
I[F] &= -\frac{1}{\kappa_0} \Bigl[ F(v_0) - \kappa_0^{-1} \dot{F}(v_0) 
+ \kappa_0^{-2} \ddot{F}(v_0) + \cdots \Bigr] e^{-\kappa_0 (v-v_0)}
\nonumber \\ & \quad \mbox{} 
+ \frac{1}{\kappa_0} \Bigl[ F(v) - \kappa_0^{-1} \dot{F}(v) 
+ \kappa_0^{-2} \ddot{F}(v) + \cdots \Bigr],  
\end{eqnarray} 
in which each term within the square brackets is smaller than the
preceding one by a factor of order $(\kappa_0 \tau)^{-1}$. With this
we arrive at 
\begin{eqnarray} 
\fl
x(v) &= \frac{1}{\kappa_0} \Bigl[ \kappa_0 x(v_0) + F(v_0) 
- \kappa_0^{-1} \dot{F}(v_0) + \kappa_0^{-2} \ddot{F}(v_0) 
+ \cdots \Bigr] e^{-\kappa_0 (v-v_0)} 
\nonumber \\ \fl & \quad \mbox{} 
- \frac{1}{\kappa_0} \Bigl[ F(v) - \kappa_0^{-1} \dot{F}(v) 
+ \kappa_0^{-2} \ddot{F}(v) + \cdots \Bigr].  
\end{eqnarray} 
At large $v$ the first set of terms decay exponentially, and $x(v)$ is
dominated by the second set of terms. A good approximation is
then $x(v) \simeq -\kappa_0^{-1} F(v)$, and $x$ is seen to decay at
the same rate as $F(v)$. This shows that our integrals are indeed 
well-behaved in the limit $v \to \infty$, and that the horizon
quantities decay at the same rate as the master functions. 

\section*{References}
\bibliography{../bib/master} 
\end{document}